\documentclass[preprints,article,accept,moreauthors,pdftex]{mdpi}

\firstpage{1} 
\pubvolume{xx}
\issuenum{1}
\articlenumber{5}
\usepackage{amsmath}
\usepackage{amssymb}
\usepackage{graphicx}
\pubyear{2019}
\copyrightyear{2019}
\history{Received: date; Accepted: date; Published: date}

\pdfoutput=1

\Title{Dissipative processes and their role in the evolution of radio galaxies}

\newcommand*{\pd}[3][]{\frac{{\partial}^{#1}#2} {\partial{#3}^{#1}}}

\Author{Manel Perucho $^{1,2}$\orcidA{}}

\AuthorNames{Manel Perucho}

\address{%
$^{1}$ \quad Departament d'Astronomia i Astrof\'{\i}sica, Universitat de Val\`encia, C/ Dr. Moliner, 50, 46100, Burjassot, Valencian Country, Spain.; manel.perucho@valencia.edu\\
$^{2}$ \quad Observatori Astron\`omic, Universitat de Val\`encia, C/ Catedr\`atic Jos\'e Beltr\'an 2, 46980, Paterna, Valencian Country, Spain.}

\abstract{Particle acceleration in relativistic jets to very high energies occurs at the expense of the dissipation of magnetic or kinetic energy. Therefore, understanding the processes that can trigger this dissipation is key to the characterization of the energy budgets and particle acceleration mechanisms at action in active galaxies. Instabilities and entrainment are two obvious candidates to trigger dissipation. On the one hand, supersonic, relativistic flows threaded by helical fields, as expected from the standard formation models of jets in supermassive black-holes, are unstable to a series of magnetohydrodynamical instabilities, such as the Kelvin-Helmholtz, current-driven, or possibly the pressure-driven instabilities. Furthermore, in the case of expanding jets, the Rayleigh-Taylor and centrifugal instabilities may also develop. With all these destabilizing processes at action, a natural question is how can some jets keep their collimated structure along hundreds of kiloparsecs. On the other hand, the interaction of the jet with stars and clouds of gas that cross the flow in their orbits around the galactic centers provides another scenario in which kinetic energy can be efficiently converted into internal energy and particles can be accelerated to non-thermal energies. In this contribution, I review the conditions under which these processes occur and their role both in jet evolution and propagation and energy dissipation.}

\keyword{Galaxies: active; Galaxies: jets; X-rays: binaries; Relativistic processes ISM: jets and outflows;  Magnetohydrodynamics; Radiation mechanisms: non-thermal}

\begin{document}


\section{Introduction}

    Relativistic jets and outflows are a common feature in radio-emitting active galaxies, associated to accretion processes onto supermassive black-holes (SMBH, \cite{eht19a,eht19b,eht19c}). According to the standard picture of relativistic jet formation, they are formed due to the extraction of rotational energy from the central hole by the magnetic field falling onto it along with the accreting gas \cite{bz77,tch11,ko12,tch15}. After formation in a very compact region of a few Schwarzschild radii ($R_s = 3\times10^{13}\,M_8\,{\rm cm}$, with $M_8$ the SMBH mass in units of $10^8\,M_\odot$), the jets propagate through up to hundreds of kiloparsecs, showing remarkable collimation at large scales in some cases, or plumed and irregular structure in others. A correlation between the total jet radio luminosity and its large-scale morphology was found by Fanaroff \& Riley \cite{fr74}, who set a morphological division of radio galaxies between those that have outer regions brighter than their centres, Fanaroff-Riley type II (FRII) radio galaxies, and those that are brighter close to the nucleus (and show decollimation at kiloparsec-scales), Fanaroff-Riley type I (FRI) radio galaxies. The former show collimation up to the interaction site with the ambient medium, where a bright radio emission region, the hotspot, is interpreted as the terminal shock that the jet particles cross. It is important to state that powerful X-ray binaries may share the formation mechanism with powerful radio sources and also present a similar morphology \cite{bk12,mar17}. 
    
     Jets can dramatically influence the evolution of the galactic environment, thus probably being a crucial piece in the evolution of galaxies and the history of star formation in the Universe. Their relativistic propagation speeds make them supersonic, which results in the generation of shocks that propagate through the interstellar medium (ISM), heating it and pushing it outwards, as revealed by the correlation between the radio-lobes and X-ray cavities that appear  at kpc-scales in radio galaxies (see, e.g., \cite{mn07,fb12,pe11,wb11,wbu12,pe14,mu16,pe17a,bi18,pe19}, and references therein).
         
    The FRI/FRII morphological dichotomy has been related with different estimates of jet kinetic power, with FRIIs corresponding to more powerful jets and FRIs corresponding to weaker jets. Although the jet kinetic power is a difficult magnitude to estimate, this seems to be a solid trend (e.g., \cite{gc01}), with FRII typical jet powers $L_j \geq 10^{45}\,{\rm erg/s}$, and $L_j \leq 10^{44}\,{\rm erg/s}$ for FRIs. At the transition powers, the ambient medium can play a role in the global morphology and this is probably the region where the hybrid morphology sources can develop, in which one of the jets shows FRI morphology, whereas the second jet is an FRII \cite{gkw00}. 
    
    The remarkable symmetry observed between jets and counter-jets at the large scales in FRIs has been interpreted and modelled as caused by strong deceleration \cite{la96,lb02a,lb02b,lb12,lb14}, as opposed to the expected jet to counter-jet brightness asymmetry in the case of relativistic flows, which is caused by relativistic Doppler boosting. This asymmetry is precisely observed in FRIIs. G. Bicknell \cite{bi84,bi86a,bi86b,bi94,bi95}, D. De Young \cite{dy86,dy93} and S. Komissarov \cite{ko88,ko90a,ko90b} set the basis of the FRI jet evolution paradigm, which is grounded on the transition from a supersonic flow to a transonic, turbulent flow, via entrainment and deceleration. Since then, a large number of works have tried to sort out what is the exact physical process that is responsible for jet deceleration in low-luminosity radio galaxies. The proposed scenarios include the development of long-wavelength ($\lambda \geq R_j$, with $R_j$ the jet radius) Kelvin-Helmholtz or kink current-driven instabilities  (KHI and CDI, respectively, \cite{ro08,pe10,tch16}), strong recollimation shocks \cite{pm07}, entrainment by stellar winds \cite{ko94,bo96,hb06}, or entrainment via turbulent mixing from the shear-layer between the jet and its surrounding medium \cite{bi84,dy93,wa09}. All these processes imply strong dissipation, but they differ in the way this occurs. Long-wavelength instabilities and strong recollimation shocks imply relatively fast dissipation, whereas entrainment caused by small-wavelength instabilities or stellar-winds is a slower process that can thus extend the deceleration and dissipation along longer distances. In \cite{kh12,lb14}, the authors used the modelling of radio images of several FRI radio galaxies to conclude that the deceleration process is extended (in all but one of the studied cases) and, moreover, coincides with a region in which the jets are X-ray bright. Altogether, this result favors the progressive dissipation scenario. 
      
  Independently of their large-scale morphology, the radio galaxies emit radiation across the whole electromagnetic spectrum, also at gamma-rays, as shown by the {\em Fermi}-Large Area Telescope (see, e.g., \cite{dg16}, and references therein). Interestingly, however, the number of detections is relatively larger for FRIs than for FRIIs, which can not be explained except if it is caused by intrinsic differences between the jets of both populations \cite{gr12,rl18}. It has been suggested that this can be due to different beaming and jet structures \cite{gr12}. Furthermore, it has been shown that the gamma-ray emission from the FRI radiogalaxy Centaurus~A is extended (not localized at the core, but in the lobes) and diffuse \cite{ab10}. Again, this requires extended energy dissipation that can be invested in particle acceleration via different mechanisms. Within the jet, both the acceleration at shearing layers \cite{rm02,so02,rd04,lo07,sa09,rd16,liu17,ki18,we18} and at the interaction site with stars or clouds (e.g., \cite{bp97,bba10,ara13,wyk13,wyk15,vi17,tab19}) have been invoked to explain high-energy (HE) and very-high-energy (VHE) from radio galaxies, although at different positions along the jet. 
  
   In this contribution, I summarize the physical processes that represent possible ways to dissipate energy in jets, namely the development of instabilities and mass entrainment. I also argue that the conversion of the initial jet kinetic energy into internal energy via shocks and turbulent dissipation at the mixing regions could favor particle acceleration. As a consequence, this could explain, at least to some extent, the diffuse HE emission detected in FRI sources, because these processes are more efficient in the case of decelerating jets.    
   
   The paper is structured as follows: In Sect. 2, I focus on the instabilities that can arise in relativistic jets, separating the linear and non-linear regime phases and I discuss the possible explanation for the stability of FRII jets as compared to FRIs. Sect. 3 is devoted to jet-star/cloud interactions. In Sect. 4 I discuss the global role of stellar mass-load in jet dynamics, and Sect. 5 includes the discussion, in which I develop on the production of high-energy and very-high-energy radiation. 
 
\section{Jet stability} \label{sec:st}

\subsection{Why are jets unstable?}

  In the same way that a kid tests if something may be broken by trying to break it, the way to know whether a system is unstable is destabilizing it. This can be easily done by introducing a small perturbation in the variables that govern it and linearizing the resulting equations. The equations that we use to model the plasma forming the relativistic jets are the relativistic magnetohydrodynamics (RMHD) equations. The stress-energy tensor for the magnetised plasma is $T^{\mu \nu} = T^{\mu \nu}_G + T^{\mu \nu}_{EM}$, with\footnote{I will use $c=1$ throughout the paper.} 
  
\begin{equation}
T^{\mu \nu}_G = \rho h u^{\mu} u^{\nu} + p g^{\mu \nu},
\end{equation}
where the Greek indices $\mu,\,\nu=0,1,2,3$ denote the four dimensions of the Minkowski space-time, $\rho$ is the flow rest-mass density, $h$ the gas specific enthalpy, $u$ is the four-velocity, $p$ is the gas pressure, and $g$ the Minkowski metric, and

\begin{equation}
T^{\mu \nu}_{EM} = F^\mu_\beta F^{\nu \beta} \,-\, \frac{1} {4} g^{\mu \nu} F_{\alpha \beta} F^{\alpha \beta},
\end{equation}
with $F^{\alpha \beta}=\epsilon^{\alpha\beta\mu\nu} b_\mu u_\nu$ the elements of the electromagnetic field tensor satisfying the Maxwell equations:
\begin{equation}
\partial_{\alpha} F_{\mu \nu} = 0 \qquad \nabla_\mu F^{\mu \nu} = -J^\nu,
\end{equation}
\noindent
where $\epsilon^{\alpha\beta\mu\nu}$ is the Levi-Civita tensor, $b_\mu$ is the magnetic field in the fluid rest-frame, and $J^\nu$ is the charge four-current. This results in the following expression for the stress-energy tensor of a magnetized, perfect fluid:
\begin{equation}
T^{\mu \nu} = \rho h^* u^\mu u^\nu + p^* g^{\mu \nu} - b^\mu b^\nu,
\end{equation}
with $h^*= h + |b|^2 /\rho$ the gas plus field enthalpy, $p^*$ the gas plus magnetic pressure ($p^*=p_{\rm gas} + |b|^2/2$), and $|b|^2 = b_\alpha b^\alpha= B^2/\gamma^2 + (\mathbf{v}\cdot\mathbf{B})^2$. Note that a $\sqrt{4\pi}$ factor is embedded into the definition of the magnetic fields (and this definition is used throughout the paper). 

The equations that dictate the flow dynamics, according to our mathematical modelling are derived from the following conservation laws:

\begin{equation}
\nabla_\mu\,T^{\mu \nu} = 0 \qquad  \nabla_\mu (\rho u^\mu) = 0.
\end{equation}  

The resulting equations (see \cite{lei05}) are the conservation of mass

\begin{equation} \label{eq:mass}
\pd{\gamma \rho} {t} + \nabla_i (\gamma \rho v^i) = 0,
\end{equation}
where $\mathbf{v}$ is the three-velocity vector (bold-faced variables express three-vectors) and $\gamma$ is the Lorentz factor. The conservation of momentum, 
 
\begin{equation} \label{eq:mom}
\pd{\gamma^2 \rho h^* v^i} {t} + \nabla_j (\gamma^2 \rho h^* v^i v^j + p^* \delta^{i j} - b^i b^j) = 0
\end{equation} 
where $i, \, j = 1,2,3$ represent the spatial coordinates, $v^i$ are the components of the three-velocity vector, and $b^i = B^i / \gamma + v^i \gamma (\mathbf{v}\cdot \mathbf{B})$ are the spatial components of the magnetic four vector, with $\mathbf{B}$ the magnetic field in the observer's frame. And, finally, the energy equation can be obtained from $\nabla_\mu T^{\mu 0} = 0$:

\begin{equation} \label{eq:ene}
\pd{(\rho h^* \gamma^2 - p^* - b^0 b^0 - \rho \gamma)} {t} + \nabla \cdot (\rho h^*\gamma^2 \mathbf{v} - b^0 \mathbf{b} - \rho \gamma \mathbf{v}) = 0,
\end{equation}
with $b^0 = \gamma (\mathbf{v} \cdot \mathbf{B})$.   

The field equations for ideal MHD are 

\begin{equation}\label{eq:field}
 \pd{\mathbf{B}} {t} + \nabla \times \mathbf{E} = 0, \qquad \nabla \cdot \mathbf{B} = 0. 
\end{equation}
where $\mathbf{E} = - \mathbf v \times \mathbf{B}$ is the electric field three-vector.

The set of equations \ref{eq:mass}-\ref{eq:field} can be applied to the description of an infinite flow, initially axisymmetric, embedded within a given ambient medium.
The next step is to introduce a perturbation in each of the physical variables in the observer's frame, $X$, and linearize the equations (in cylindrical coordinates): 

\begin{align*}
X(r) & \longrightarrow X_{0}(r) + X_{1}(r,\phi,z,t),
\end{align*}
with $X_0(r)$ the equilibrium distribution of the variable $X$ (with $X=\mathbf{v}, \, \rho, \, p, \, \mathbf{B}$) and $X_1$ its perturbation. The resulting set of linearized equations (see, e.g., \cite{har07}) that result after elimination of the terms that reproduce the equilibrium configuration and higher-order terms are 

\begin{equation}
\pd{}{t}\left[\gamma_0 \rho_1 + \gamma_1 \rho_0\right] \,+\,\mathbf{\nabla}\cdot\left[\gamma_0 \rho_1 \mathbf{v} + \gamma_0 \rho_0 \mathbf{v}_1 + \gamma_1 \rho_0 \mathbf{v} \right]= 0,  
\end{equation}
for the continuity equation,

\begin{equation}
\begin{split}
\pd{}{t} \left[ \gamma_{0}^2 (\rho h)_1 - P_1 + 2 \gamma_0^4 \left(\mathbf{v}\cdot\mathbf{v}_1\right) (\rho h)_0 \right] + \nabla \cdot \left[\gamma_0^2 (\rho h)_1 \mathbf{v} + 2\gamma_0^4 \left(\mathbf{v}\cdot\mathbf{v}_1\right) (\rho h)_0 \mathbf{v} + \gamma_0^2 (\rho h)_0 \mathbf{v}_1\right] \\
+ \pd{}{t} \left[B_0^2 (\mathbf{v}\cdot\mathbf{v}_1) + (1+v^2)\mathbf{B}_0\cdot\mathbf{B}_1 - (\mathbf{v}\cdot\mathbf{B}_1 + \mathbf{v}_1 \cdot \mathbf{B}_0) \mathbf{v}\cdot\mathbf{B}_0/c\right]\\
+ \nabla\cdot\left[2(\mathbf{B}_0\cdot\mathbf{B}_1)\mathbf{v} + B_0^2\mathbf{v}_1-(\mathbf{v}\cdot\mathbf{B}_0)\mathbf{B}_1 - (\mathbf{v}\cdot\mathbf{B}_1)\mathbf{B}_0 - (\mathbf{v}_1\cdot\mathbf{B}_0)\mathbf{B}_0\right] = 0.
\end{split}
\end{equation}
for the energy, and 

\begin{equation}
\begin{split}
\gamma_{0}^{2} (\rho h)_{0} \left[ \pd{\mathbf{v}_1}{t} + (\mathbf{v} \mathbf{\nabla}) \mathbf{v}_{1} \right] +\mathbf{\nabla} P_1 + \frac{\mathbf{v}} {c^2} \pd{P_{1}}{t} - (\mathbf{j}_0\times \mathbf{B}_1) - (\mathbf{j}_1\times \mathbf{B}_0) = 0,
\end{split}
\end{equation}
for the momentum. The linearized Maxwell equations are

\begin{equation}
\begin{split}
\nabla\cdot \mathbf{B}_1 = 0\\
\nabla\times\mathbf{E}_1 = -\pd{\mathbf{B}_1} {t}, 
\end{split}
\end{equation}
with $\mathbf{E}_1 = - \mathbf{v}_0\times \mathbf{B}_1 - \mathbf{v}_1\times \mathbf{B}_0$.


Before solving these equations, different approximations/simplifications can be made (and have been made by different authors) in order to study jet stability. The relevance of this selection lies on the fact that different types of instability arise from the different terms, which leads to a selected study of a given instability whereas those embedded in the neglected terms are obviated. According to \cite{bo16}, where the non-relativistic regime was studied, the different terms arising from the linearized equations, in the cold limit, lead to the complete list of instabilities that can arise in jetted flows, namely, the KHI, CDI, centrifugal buoyancy (Parker instability) and magnetorotational instability (MRI, see also \cite{ha00}).\footnote{The short wavelength pressure-driven (or Z-pinch) instability, which can grow for a particular initial distribution of magnetic field and gas pressure \cite{beg98} can be taken as a particular type of CDI.}

  The instabilities that can arise in relativistic outflows have a range of causes: 
  \begin{itemize}
    	\item KHI arises due to transition layers (either contact discontinuities or continuous changes) in the velocity component tangential to this layer; 
	\item Rayleigh-Taylor instability (RTI) can grow in expanding/contracting jets, with the equivalent to the gravity acceleration being given by the changes in the radial velocity; 
	 \item centrifugal instability (CFI) also grows in expanding jets with non-zero azimuthal velocities; 
	 \item CDI arises due to deformations in the toroidal component of the magnetic field, and the so-called pressure-driven instability arises when the magnetic rings that constitute the magnetic structure in the case of toroidal component dominated field are displaced from their equilibrium configuration (under certain magnetic field radial distributions, \cite{beg98}); 
	 \item centrifugal buoyancy is caused by rotation, when the centrifugal force is large enough to bend the magnetic lines; 
	 \item MRI can develop in a magnetized, rotating jet with differential rotation (and decreasing angular velocity with radius; see \cite{bo16}). 
	 \end{itemize}
	 
	 The growth of these instabilities to the nonlinear regime could imply strong dissipation of kinetic energy, and also magnetic energy via reconnection of magnetic lines (thus departing from ideal MHD).
	 
    In the following list I summarize the different approximations that have been studied in the case of relativistic flows, adding some of the relevant references in which the priority is given to those where the focus is on the linear regime:\footnote{There are aspects relevant to the parameter space covered by each of the referred papers (e.g., magnetically versus particle dominated, or hot versus cold jets...) that are not taken into account in this list.}
  
   Non-magnetized jets ($\mathbf{B}=0$):
     \begin{itemize}
         \item No rotation ($\mathbf{v} = v^z \mathbf{e}_z$): \\
              - No expansion ($\partial{v^r}/\partial{t} = 0$): KHI (e.g., \cite{ts76,bp76,fe78,har79,fe82,bi91,har00,pe04a}).\\
              - Expansion ($\partial{v^r}/\partial{t} \neq 0)$: RTI (e.g., \cite{mm13,ma17,to17}).\\
   
         \item Rotation ($\mathbf{v} = v^\phi \mathbf{e}_\phi + v^z \mathbf{e}_z$): \\
                - No expansion ($\partial{v^r}/\partial{t} = 0$): KHI (e.g., \cite{mk07,mk09}).\\
                - Expansion ($\partial{v^r}/\partial{t} \neq 0)$: CFI (e.g., \cite{gk18b}).
       \end{itemize}
   
    Magnetized jets (all non-expanding cases):
        \begin{itemize}
            \item No rotation ($\mathbf{v} = v^z \mathbf{e}_z$): \\
                  - KHI (e.g., \cite{fe80,fe81,har07}).\\
                  - CDI + KHI (e.g., \cite{os08,bo13,kim17,kim18}).\\
                  - CDI/pressure-driven instability (e.g., \cite{ip94,ip96,beg98}).\\

           \item Rotation($\mathbf{v} = v^\phi \mathbf{e}_\phi + v^z \mathbf{e}_z$): \\
                  - Centrifugal buoyancy (\cite{bo19}).\\
	     \end{itemize} 
     
   An initial equilibrium state consistent with the assumptions made has to be considered previous to solving the equations. The equilibrium state requires constant gas pressure alone if the magnetic fields and azimuthal velocity are neglected. The inclusion of the latter terms introduces complexity in the initial radial configuration of the variables due to centrifugal motion in the case of azimuthal velocity and also magnetic tension in the case of the field. These have to be compensated by specific distributions of gas pressure that stabilize the radial structure of the jet, which requires solving the transversal equilibrium equation for magnetized, relativistic jets (see, e.g., \cite{lyu99,gou12,ma15,ma16}). In the case in which expansion is considered, equilibrium is applied only at the contact discontinuity (e.g., \cite{mm13,ma17}).
   
    The solutions for the perturbations of the variables are taken to have a wave-like form $X_1(r,\phi,z,t) = X_1(r) \exp(i (\omega t \pm m\phi - k_z z))$, where the plus or minus sign that appear with the azimuthal wave number indicate the sense of the wave in the azimuthal direction. The different types of solutions are the result of the combination of the azimuthal and radial structure of the modes: 
    \begin{itemize}
     	\item If $m=0$, there is no azimuthal dependence and the instability produces axisymmetric oscillations of the jets; $m=1$ produces antisymmetric distributions of the physical parameters that result in helical oscillations; $m=2$ produces elliptical deformations, etc.       
	\item A radial structure with no zeros between the jet axis and its surface (i.e., global changes that do not cross the equilibrium value $X_0$ at any point) corresponds to the {\em surface} or {\em fundamental} modes, whereas if the radial structure crosses the equilibrium value, the modes are called {\em body} or {\em reflection} modes, with the number of zeros giving the order (one zero for the first body/reflection mode, and so on). 
      \end{itemize}
       
    Once the equilibrium is defined and thus the radial profiles of the variables are established, the system can be solved either as a dispersion relation given by matching conditions at a discontinuity between the jet and its environment, or, if the transition is continuous, solving the resulting differential equations by setting boundary conditions on the axis and at $r\rightarrow \infty$\footnote{Although there are other options: \citeauthor{ip94} (\citeyear{ip94}, \citeyear{ip96}) consider variations of the field up to the jet radius and set the boundary condition at that point}. 
    
    In the case of sheared jets, when considering a purely hydrodynamical jet, a single second-order differential equation can be derived for the pressure perturbation (e.g., \cite{bi91,ur02,pe05,pe07}). However, when a toroidal component of the magnetic field is considered, the system can only be reduced down to a couple of differential equations, typically taken for the radial velocity and the total pressure ($p^*$) perturbations. In both cases, the {\em shooting method} plus a complex-root finder (e.g, the M\"uller method, \cite{pe07}) are used. The derivation of these systems of equations can be lengthy and it is out of the scope of this review. We refer the interested reader to, e.g., \cite{bo13,kim17} for explicit derivations. 
    
    The solutions of the stability problem are usually given either taking real wavenumbers and complex frequencies (temporal approach), or complex wavenumbers and real frequencies (spatial approach). On the one hand, in the temporal approach, the instability is taken to grow in time at a given location; the imaginary part of the frequency ($\omega_i$) represents the inverse of the time needed by the instability to e-fold its amplitude and it is called {\em growth-rate}. On the other hand, in the spatial approach, the imaginary part of the wavenumber ($k_i$) represents the inverse of the distance it takes for the instability to e-fold in amplitude. The inverse of $k_i$ is named growth-length or growth-distance. In the rest of this paper, I will typically refer to growth-rates, but the reader has to keep in mind that there is an inverse relation between this quantity and growth-lengths, so that an increase of the growth-rate would imply a reduction of the length required to multiply the instability amplitude by a factor $e$. Although this relation is only qualitative and the quantitative conversion from the temporal to the spatial approach requires paying close attention to the group velocity ($\partial \omega_R/ \partial k$, \cite{har86}), the qualitative relation is sufficient for the purposes of this paper.

    Independently of the type of instability that develops and the jet initial properties, and of the fact that the jet is disrupted or not, the instabilities dissipate kinetic or magnetic energy and can thus act as efficient particle-acceleration mechanisms (e.g., \cite{fe79,mi16}). 
   
\subsection{The linear regime}  \label{sec:lin} 
Out of all the aforementioned papers, a number of clear conclusions can be extracted regarding the linear growth of jet instabilities. In this subsection, I summarize these ideas.
 
\subsubsection{Hydrodynamical jets}   
   Relativistic hydrodynamic ($\mathbf{B}\rightarrow0$) jets are KH unstable because they are typically supersonic and present a velocity transition (either if it is modelled as a contact discontinuity or as a shear-layer) with the ambient medium (e.g., \cite{ts76,bp76,har79,fe78,fe80,fe81,bi91}). 
   
\begin{figure}
\centering
\includegraphics[trim={1cm 0.9cm 6.75cm 17.5cm},clip,width=0.45\textwidth]{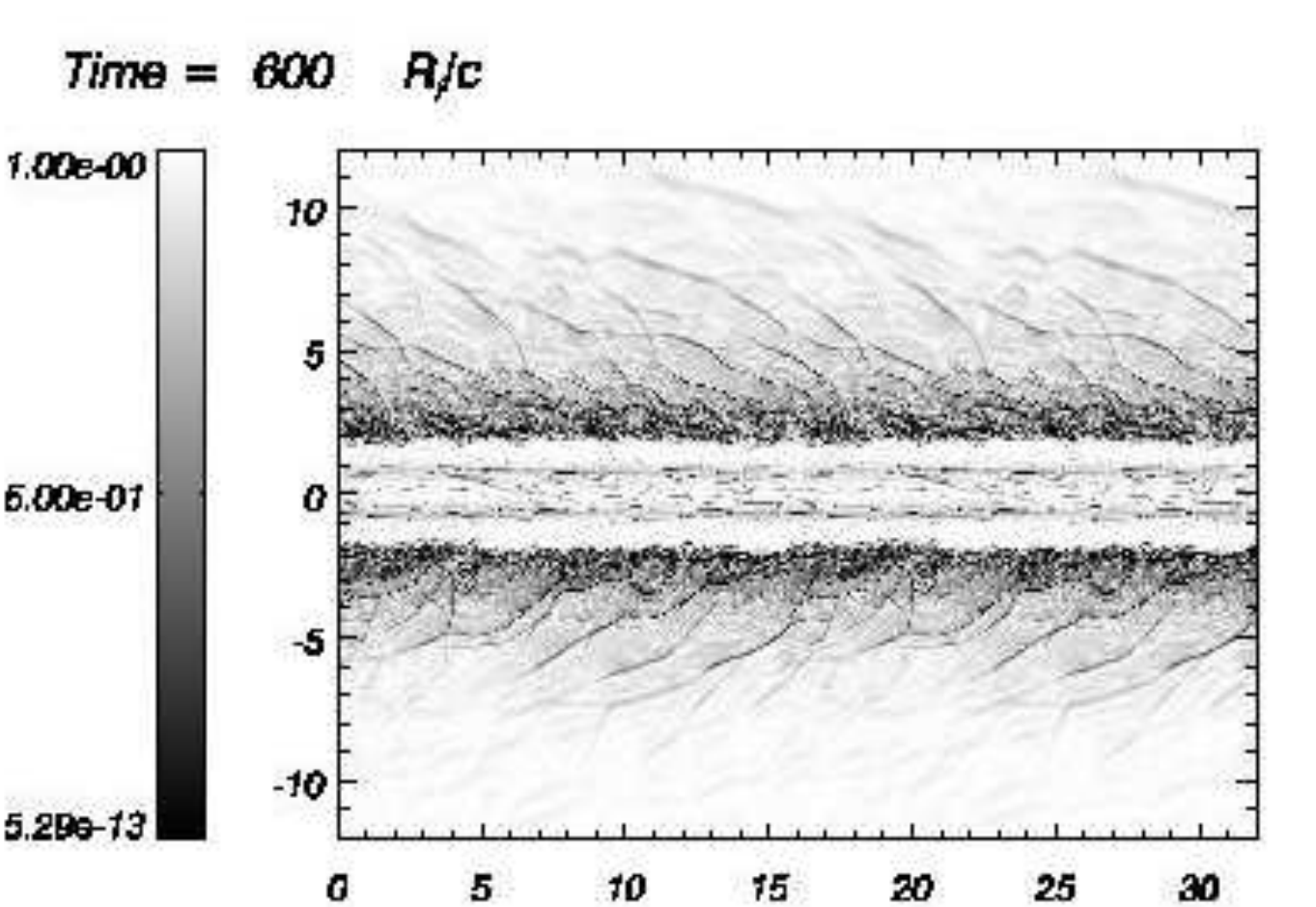} \,\,\quad\,\,\includegraphics[trim={1cm 0.9cm 6.75cm 17.5cm},clip,width=0.45\textwidth]{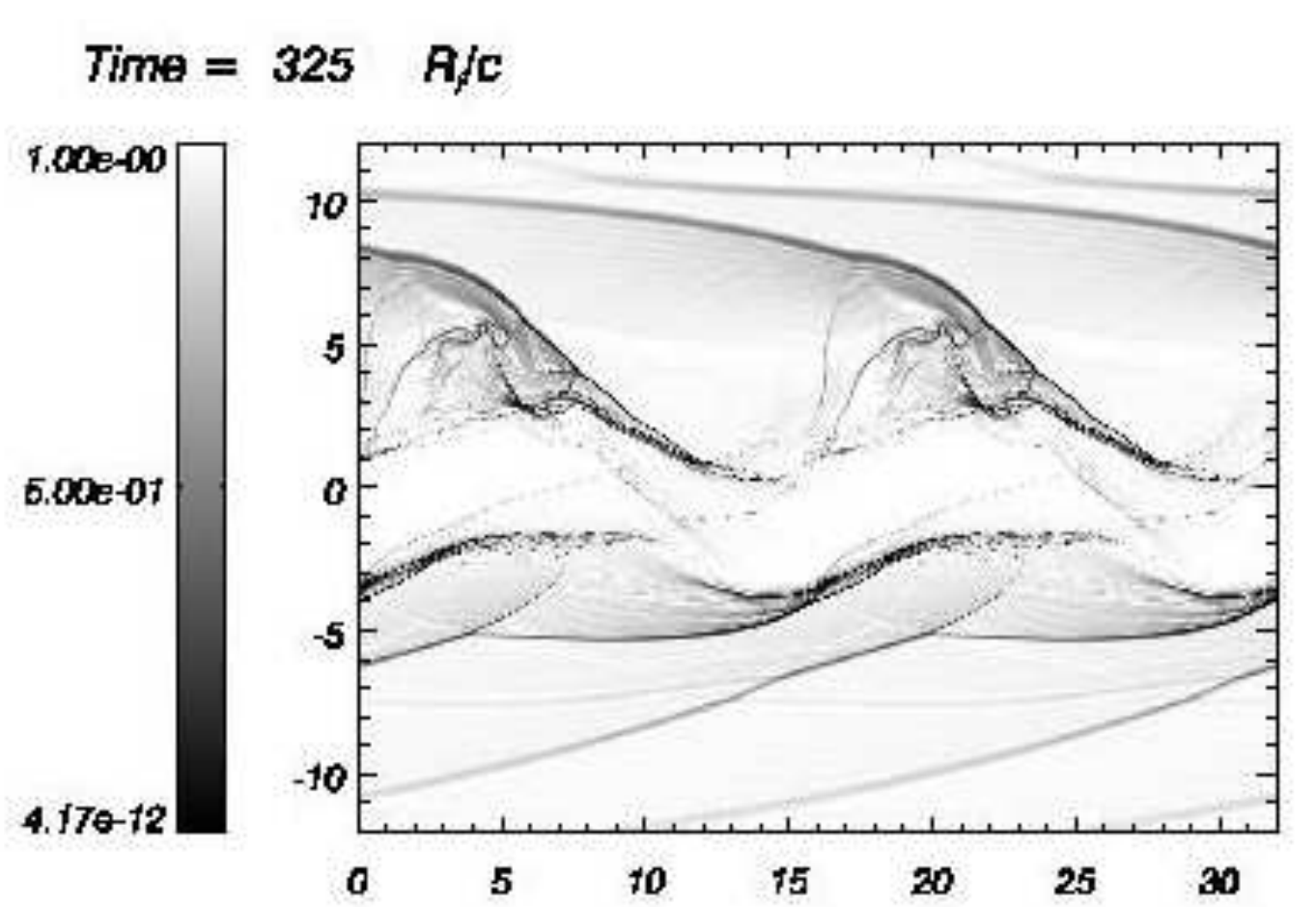}
 \includegraphics[trim={1cm 0.9cm 6.75cm 17.5cm},clip,width=0.45\textwidth]{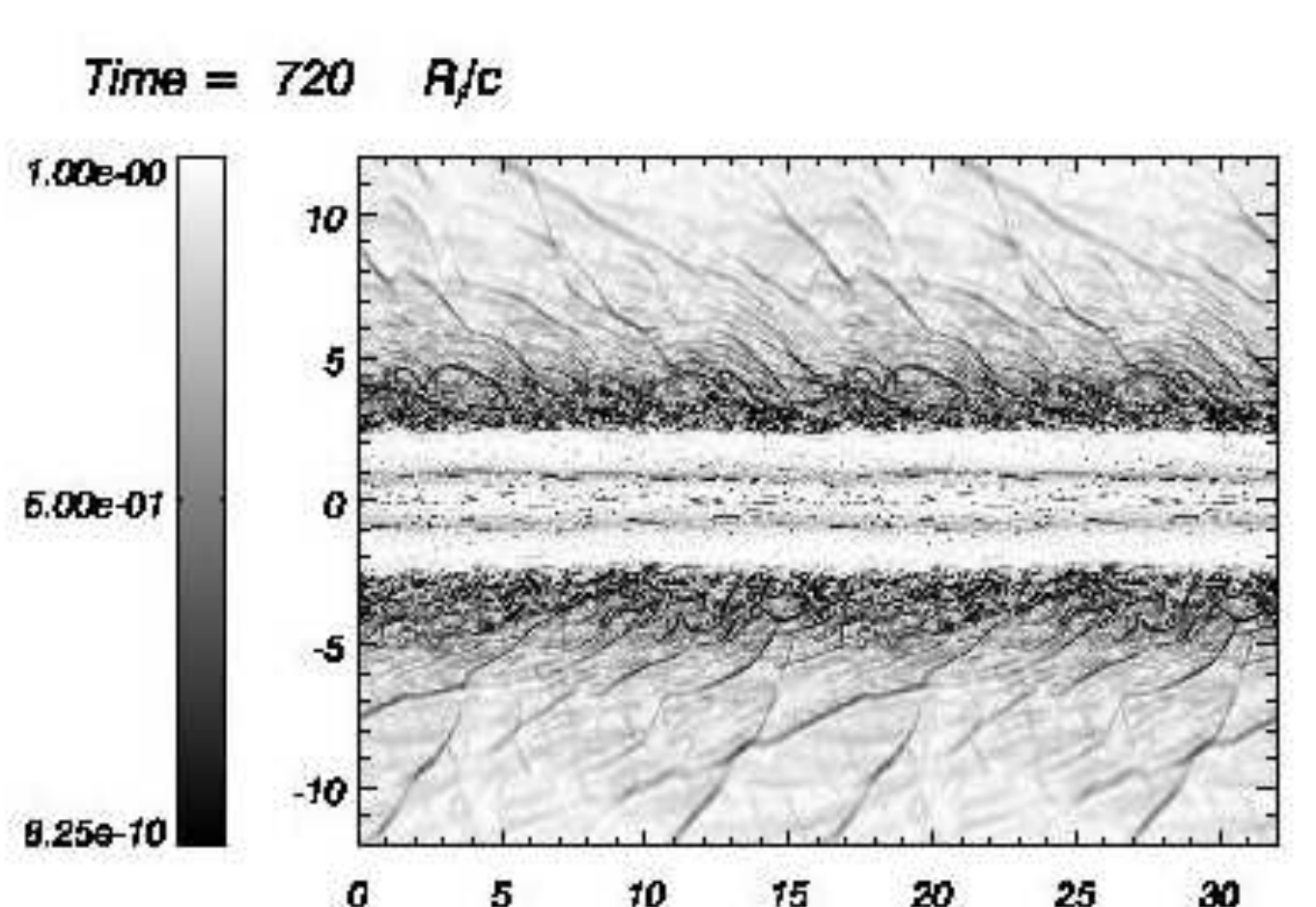} \,\,\quad\,\,\includegraphics[trim={1cm 0.9cm 6.75cm 17.5cm},clip,width=0.45\textwidth]{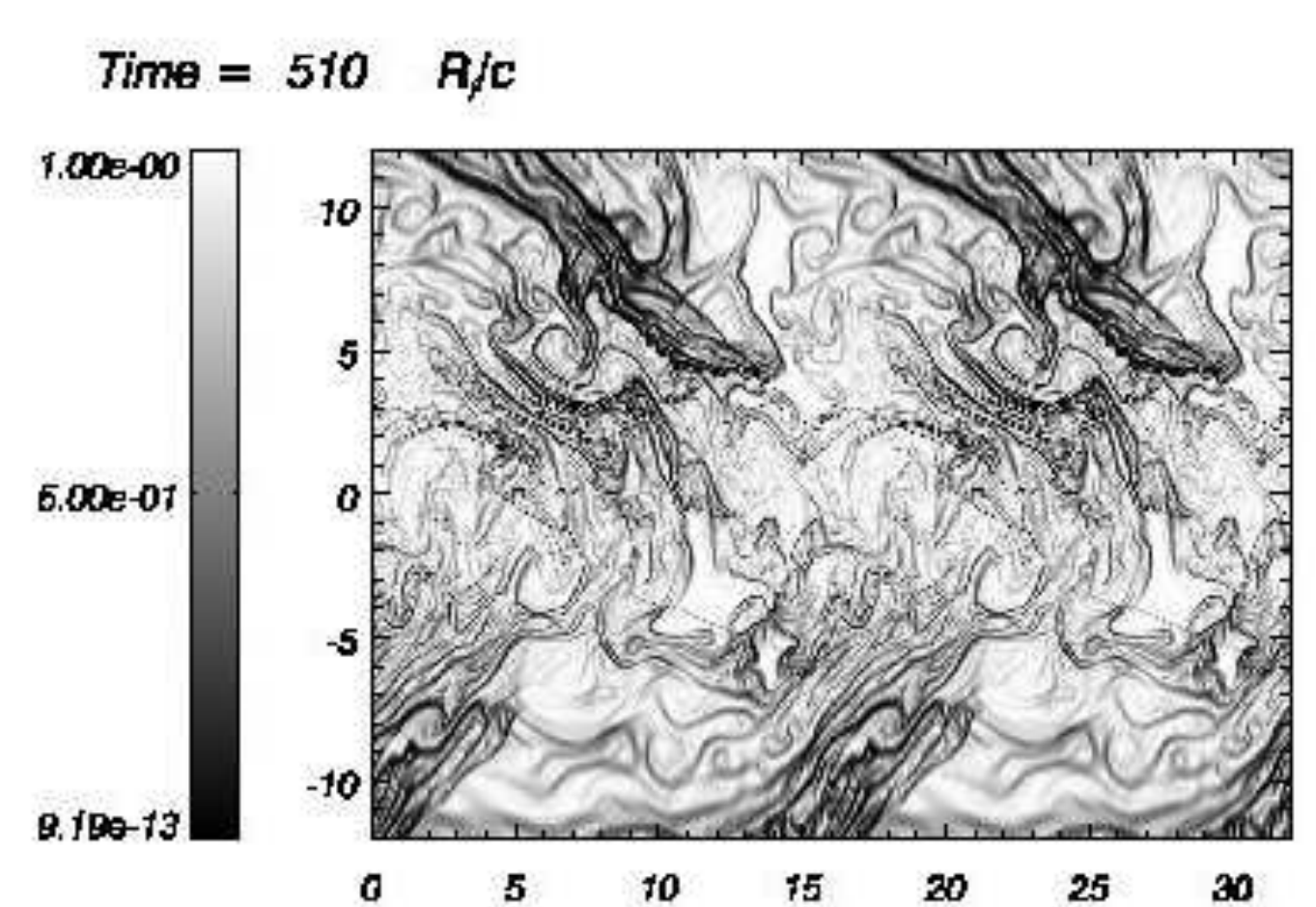}
\caption{Schlieren plots (enhanced rest-mass density gradients showing waves and shocks) from numerical simulations of two jets at different times \cite{pe07}. The left column shows the evolution of a jet (fast and relatively cold) in which resonant modes develop at the shear-layer; as a result of the strong dissipation caused by these modes, a thick, hot layer develops surrounding the collimated flow that preserves basically unchanged properties \cite{pe05}. The right column shows the evolution of a slower jet, in which resonant modes do not dominate the growth rates at the linear regime; in this case, the jet is disrupted by the long-wavelength helical mode.}\label{fig1}
\end{figure}

  The growth rates of the KHI modes depend basically on the jet Lorentz factor, internal energy density, transversal structure (sheared jets), and jet expansion (see, e.g., \citeauthor{har79} \cite{har84,har86,har87,har00}, and references therein; \citeauthor{pe04a} \cite{pe04a,pe05,pe07}). Typically, slower (smaller Lorentz factor) and hotter (larger internal energy density) jets show larger values of the growth-rates \cite{pe04a,pe05}. In addition, the introduction of a shear layer contributes to a reduction of the growth rates of modes with small wavelengths (comparable to or smaller than the width of the layer), but for the case of faster jets, in which fast-growing, short-wavelength resonant modes appear \cite{ur02,pe05,pe07}.  Interestingly, the resonant modes reported in \cite{pe07} grow fast enough in the shear region to generate a wide, hot shear-layer around the jet spine, which remains basically untouched (see Figure~\ref{fig1}). In the next section, I will discuss the implications of this and other post-linear effects of unstable modes in relation to long-term jet stability. Furthermore, jet expansion favours an increase of the mode wavelength with distance and this typically implies a reduction of the KHI growth-rates \cite{har86,pk15}.

   The Lorentz factor plays a dual role: In addition to smaller growth rates, the effective distances over which the modes grow are also longer for higher Lorentz factors ($\geq 10$) \cite{pe04a,pe05}. This is the first important stabilizing factor in relativistic jets when compared to sub-relativistic ones. In contrast, low Lorentz factor jets present higher growth-rates and are thus more prone to the full development of the instability to the nonlinear regime, mass-load and deceleration within relatively short length-scales.
   
\begin{figure}
\centering
\includegraphics[trim={0cm 0cm 0cm 5cm},clip,width=0.6\textwidth]{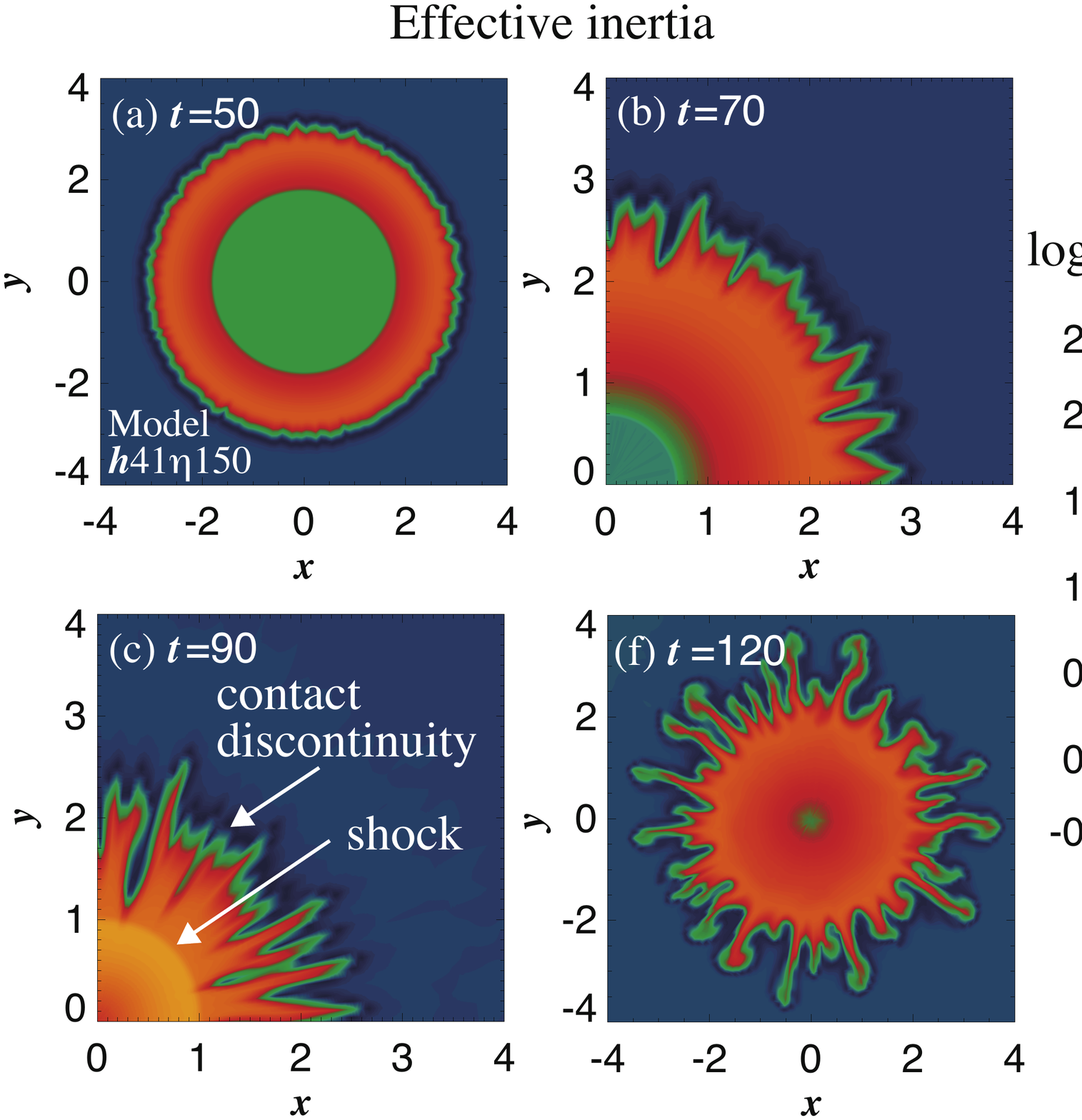}
\caption{The image shows panels of the logarithm of the effective inertia ($\gamma^2 \rho h$ in units of ambient rest-mass energy density, $\rho_a\, c^2$) along an expanding and recollimating jet from the simulations in \cite{mm13}. The development of the RTI can be clearly observed (adapted from \cite{mm13}, courtesy of Jin Matsumoto).}\label{fig2}
\end{figure}

   The requirement of small-scale instabilities that can explain the way in which FRI jets decelerate along the first kiloparsecs \cite{lb14} has brought the attention of theorists upon the possibility that the RTI or the CFI develop in expanding jets. Different authors have studied the development of Rayleigh-Taylor instability in jets undergoing expansion and recollimation (\cite{mm13,ma17,to17}, see Fig.~\ref{fig2}). In particular, in reference \cite{ma17}, the authors develop the linear analysis and confirm the prediction given by \cite{mm13} about the stability condition
   
   \begin{equation} \label{eq:rti}
   \frac{\rho_1 h_1' \gamma_1^2} {\rho_2 h_2' \gamma_2^2} > 1,
   \end{equation}
   where subscripts $1$ and $2$ indicate the jet flow and the cocoon, respectively, $\rho$ is the rest-mass density, $\gamma$ is the Lorentz factor and $h'$ is defined as
   
   \begin{equation}
   h' := 1 +\frac{\Gamma^2} {\Gamma-1}\frac{p} {\rho}, 
   \end{equation} 
   with $\Gamma$ the adiabatic exponent of the fluids, assumed to be constant across the contact discontinuity. Equation \ref{eq:rti} states that the system is unstable when the inertia of the jet is larger than that of the cocoon or surrounding medium. The authors concluded that this instability is thus mainly excited in oscillating jets with high inertia, as opposed to KHI, where growth-rates are reduced by the jet inertia. In \cite{to17}, the authors have shown that the development of the RTI instability (or Richtmeyer-Meshkov instability in the case of impulsive expansion) can be extended to different spine-sheath configurations. This work shows that the spine does not necessarily need to be 'heavier' than the cocoon for the RTI to develop, because the acceleration vector is inverted in the reconfinement region with respect to the expansion region and the roles of the jet and cocoon are thus interchanged, also producing small-scale mixing at the jet boundary.
   
\begin{figure}
\centering
\includegraphics[width=0.7\textwidth]{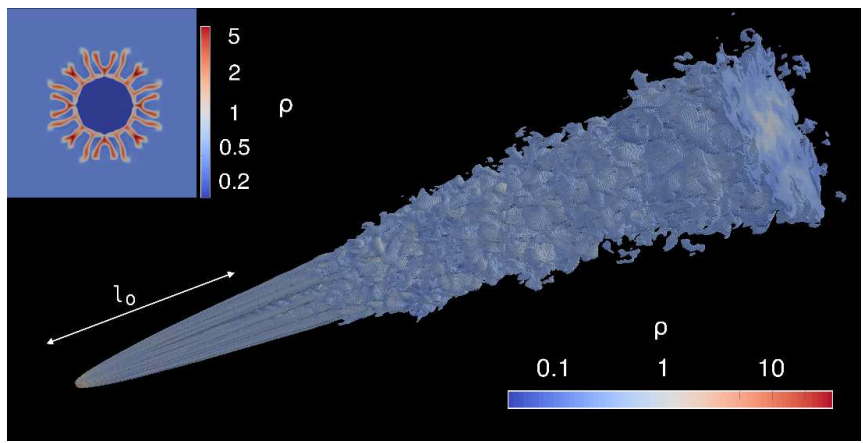}
\caption{Rest-mass density contour (in arbitrary units) from a simulation in which the development of RTI produces the decollimation and deceleration of a relativistic jet \cite{gk18a} (courtesy of Konstantinos Gourgouliatos).}\label{fig3}
\end{figure}

   \citeauthor{mk07} \cite{mk07,mk09} studied the development of instabilities in the shear between rotating cylinders, and interpreted them as RTI developing due to centrifugal acceleration. \citeauthor{gk18b} \cite{gk18b} have shown that this is actually the relativistic version of the CFI and that it may grow in flows with curved streamlines such as jets with oscillatory cross-sections caused by expansion and recollimation (see Fig.~\ref{fig3}). The authors generalize the Rayleigh criterion to the relativistic case for a discontinuity separating two rotating flows to
   \begin{equation} \label{eq:cfi}
    \Psi_2 - \Psi_1 < 0,
   \end{equation}
with 
\begin{equation} 
\Psi = \rho h \gamma^2 (\Omega R^2)^2,  
\end{equation}       
where $\Omega$ is the angular velocity, for a ring of fluid pushed from medium 1 to medium 2 (which can be identified with the jet and the cocoon, as above, in the expansion phase). The authors point out that when $\Omega$ is constant across the discontinuity, this condition results in 
\begin{equation}
\rho_2 h_2 \gamma_2^2 - \rho_1 h_1 \gamma_1^2 < 0,  
\end{equation}
or $\rho_2 - \rho_1 < 0$ in the non-relativistic case, which is the RTI criterion given in Eq.~\ref{eq:rti}.

In \cite{gk18b}, the authors also derive a criterion for the case of a continuous transition between the two flows (shear). In this case, they derive
\begin{equation}
\frac{{\rm d\, ln}\Psi} {{\rm d\, ln}R} < M^2,
\end{equation}  
with $M = \gamma \Omega R/(\gamma_s c_s)$, and $\gamma_s$ the Lorentz factor of the sound speed. This criterion is, however, still to be tested by numerical simulations.    
   
\subsubsection{Magnetized jets}   
 
\paragraph{\emph KHI and CDI} 
 
    The study of the stability of relativistic jets threaded by magnetic fields started from the relatively easier configurations of purely axial magnetic field (e.g., \cite{har07,os08}). In reference \cite{har07}, P. Hardee studied the stability of magnetized jets with a spine-sheath structure\footnote{Despite this structure is arbitrarilly introduced in the set-up of many simulations and stability analysis, a sheath is probably surrounding the fast jet spines, either from the jet formation region, where a wind can be formed from the accretion disk \cite{bp82}, or because of simple kinetic energy dissipation at the jet-environment shearing layer.} and showed that this configuration results in the stabilization of the flow with respect to 'naked' jets, to the extent that trans-Alfv\'enic and super-Alfv\'enic spines can become stable to the KH instability (unlike in the relativistic hydrodynamics, RHD, case, where transonic spine-sheath configurations are still unstable), but for the fundamental pinch mode. The development of the KH instability in planar flows with equal velocity modulus and opposite sign, threaded by a uniform magnetic field parallel to the velocity direction was studied in \cite{os08}, where the conclusion that stronger magnetic fields and flow velocities stabilize this configuration was also reached. 
    
    Bodo and collaborators \cite{bo13} tackle the stability of relativistic, zero pressure (cold), magnetized jets with different rotation and helical field configurations leading to different current distributions within the jet, and return currents located outside the shear layer. They solve the linearized system of equations by means of the shooting method with a complex secant root finder, setting appropriate boundary conditions close to the jet axis and at radius $r \gg R_j$. A critical point that is stressed in these works is the role played by the so-called {\em resonant surfaces}, where $\mathbf{k \cdot B} = 0$: the absence of the magnetic tension in the plane perpendicular to the field line facilitates the development of the CDI in that direction. Expressing this scalar product in terms of the {\em pitch parameter} (defined as $P=r B^z/B^\phi$),\footnote{Note that the definition of the pitch varies among papers and that it can also be found as $\psi(r) = \arctan\left(B^\phi(r) / B^z(r)\right)$.} they obtain (see \cite{bo13})
   
   \begin{equation} \label{eq:res}
     \mathbf{k \cdot B}\, = \, k_z B^z\,+\, \frac{m} {r} B^\phi \,=\, 0, \qquad {\rm or}\qquad k_z P + m = 0. 
   \end{equation} 

\begin{figure}
\begin{flushleft}
\hspace{0.5 cm}
\includegraphics[trim=4cm 7.5cm 10cm 10.5cm,width=0.2\textwidth]{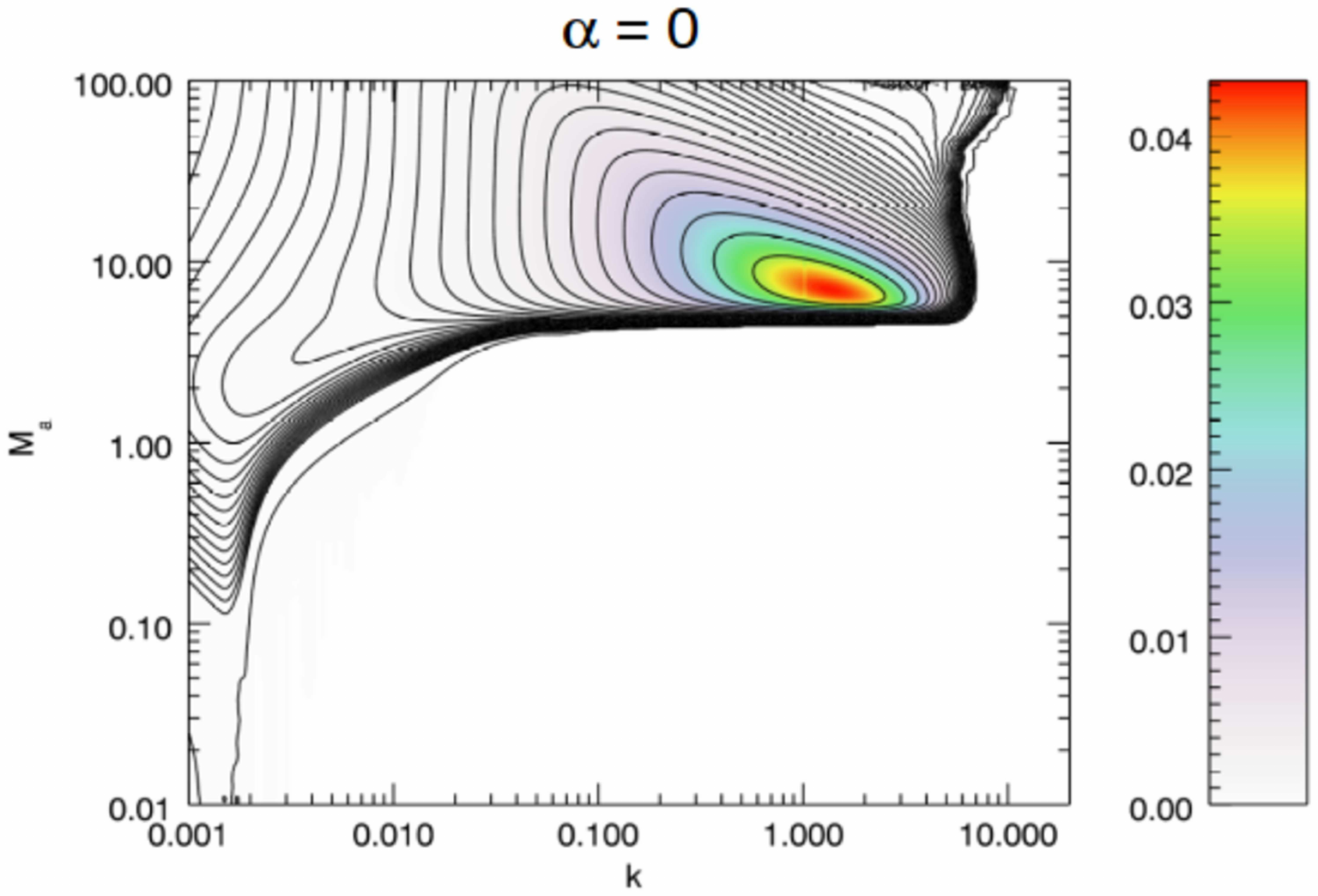} \hspace{4.5cm}\includegraphics[trim=4cm 7.5cm 10cm 10.5cm,width=0.2\textwidth]{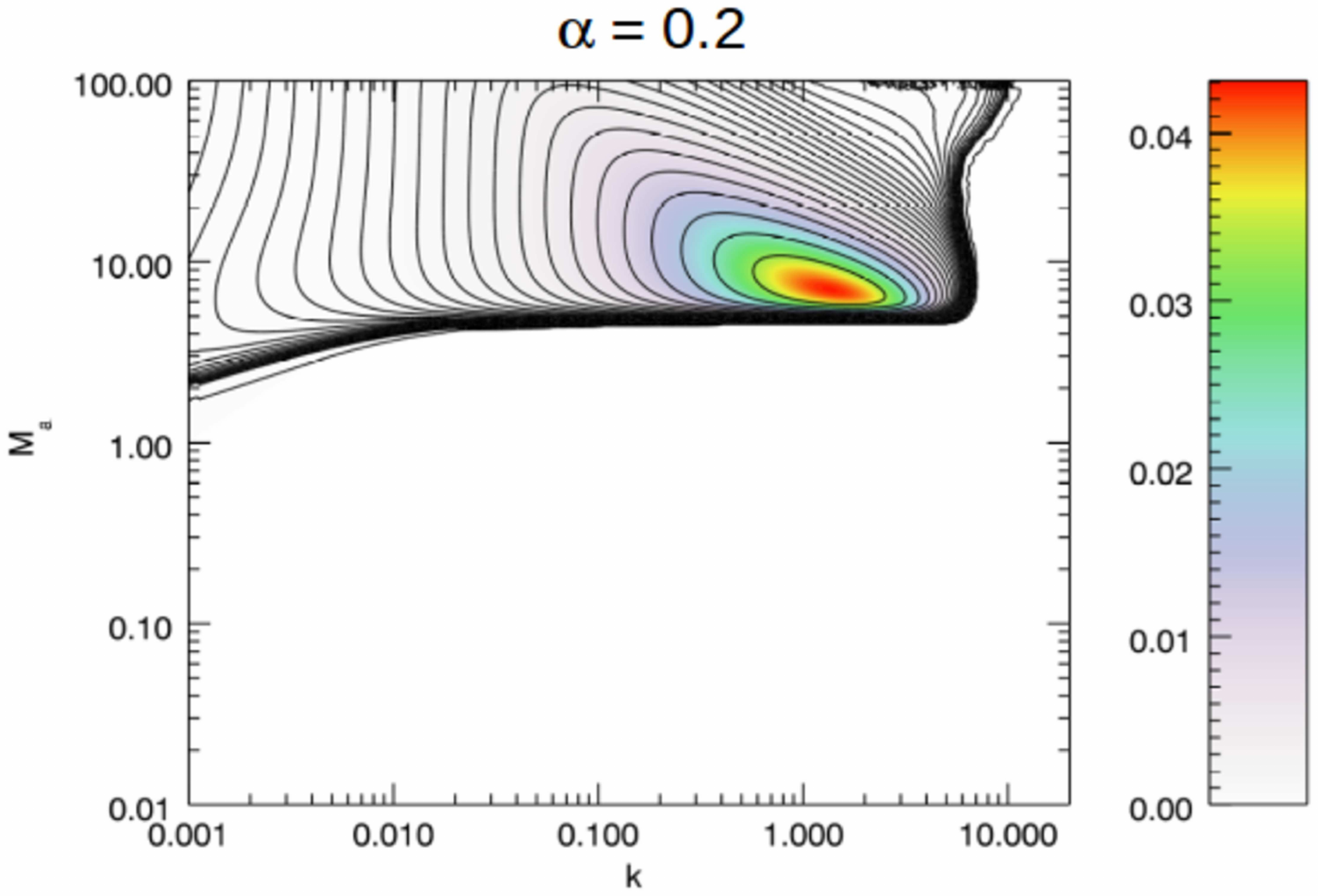}\\
\hspace{0.5 cm}
 \includegraphics[trim=4cm 7.5cm 10cm 10.5cm,width=0.2\textwidth]{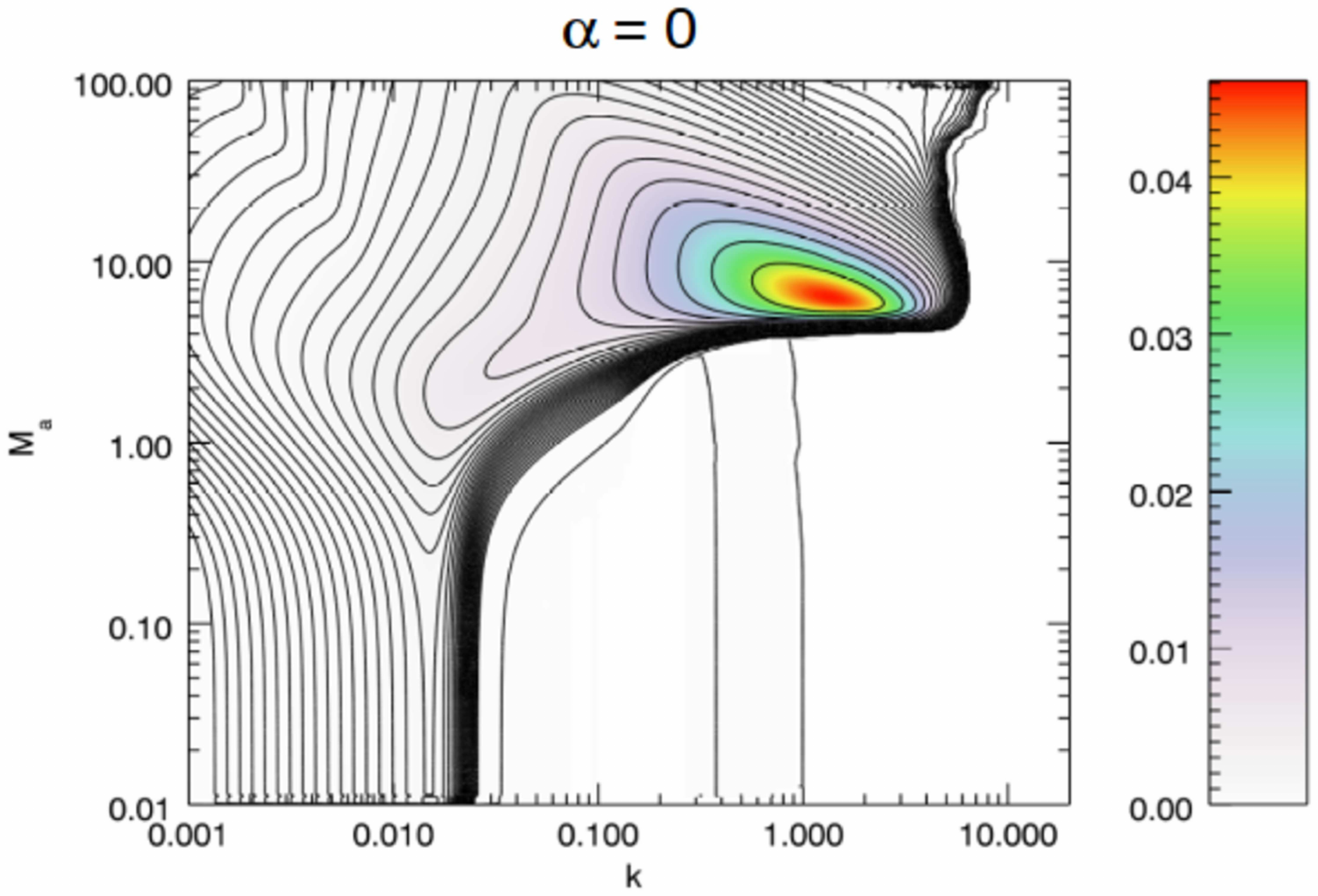} \hspace{4.5cm}\includegraphics[trim=4cm 7.5cm 10cm 10.5cm,width=0.2\textwidth]{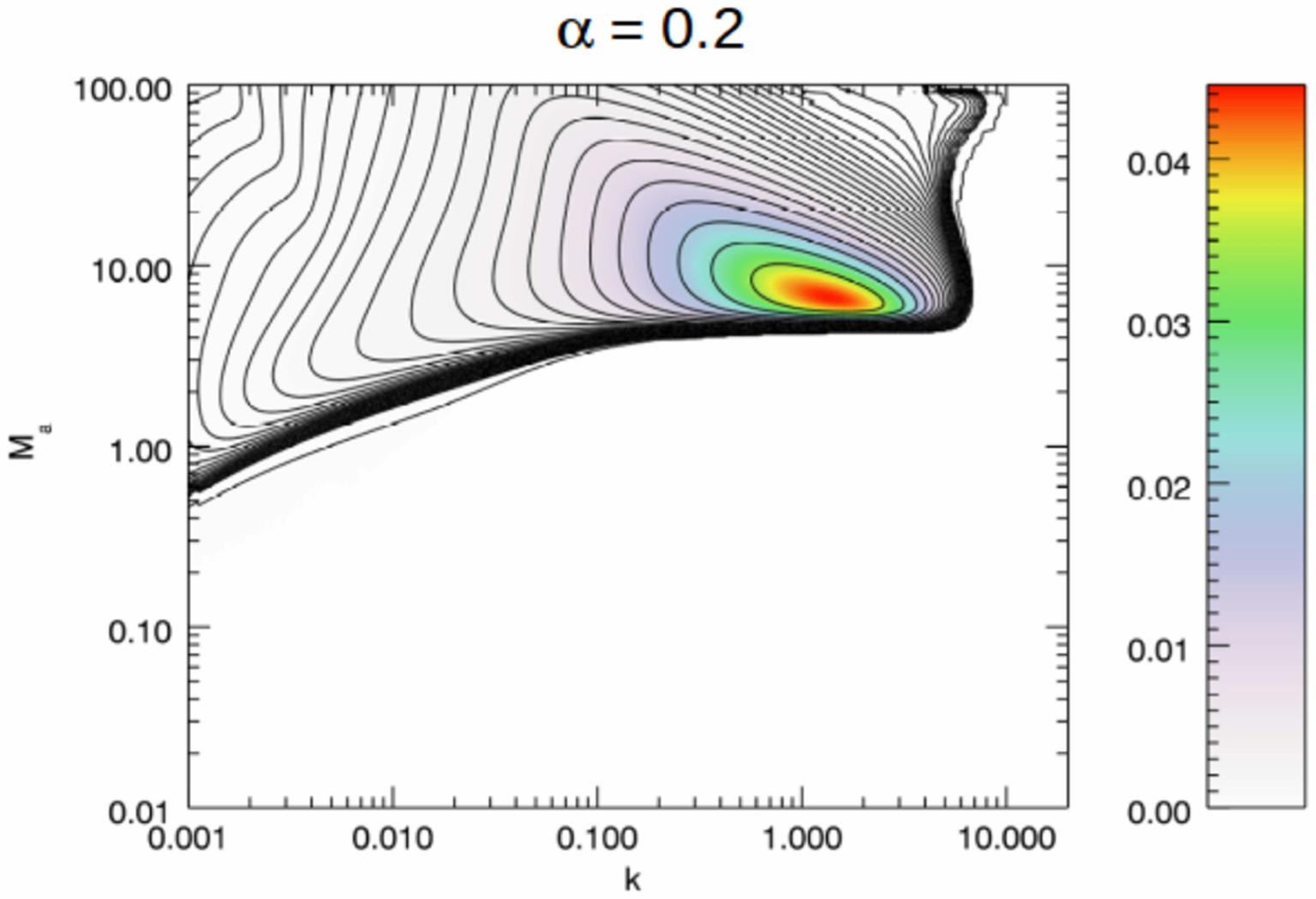}\\
 \hspace{0.5 cm}
  \includegraphics[trim=4cm 7.5cm 10cm 10.5cm,width=0.2\textwidth]{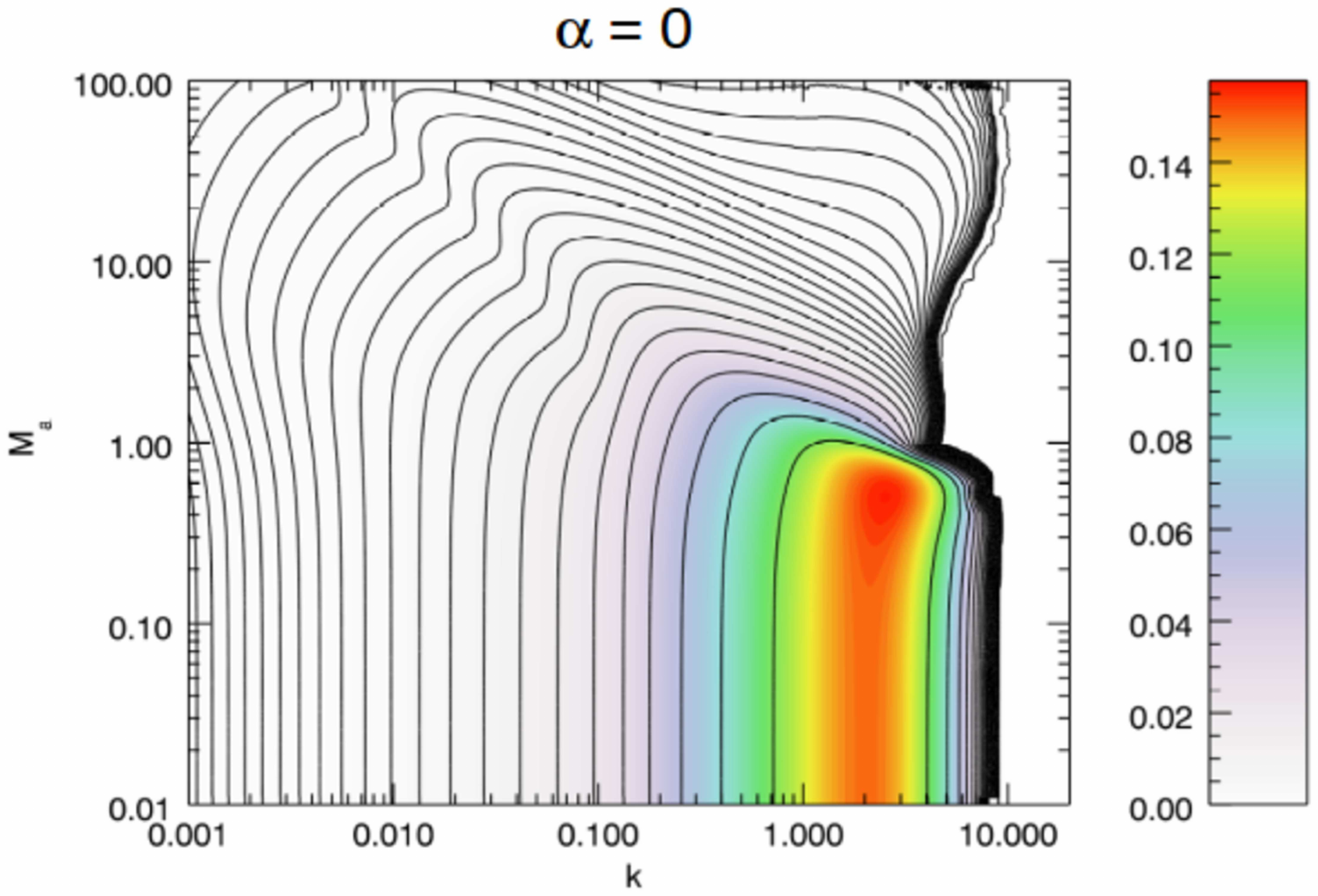}  \hspace{4.5cm} \includegraphics[trim=4cm 7.5cm 10cm 10.5cm,width=0.2\textwidth]{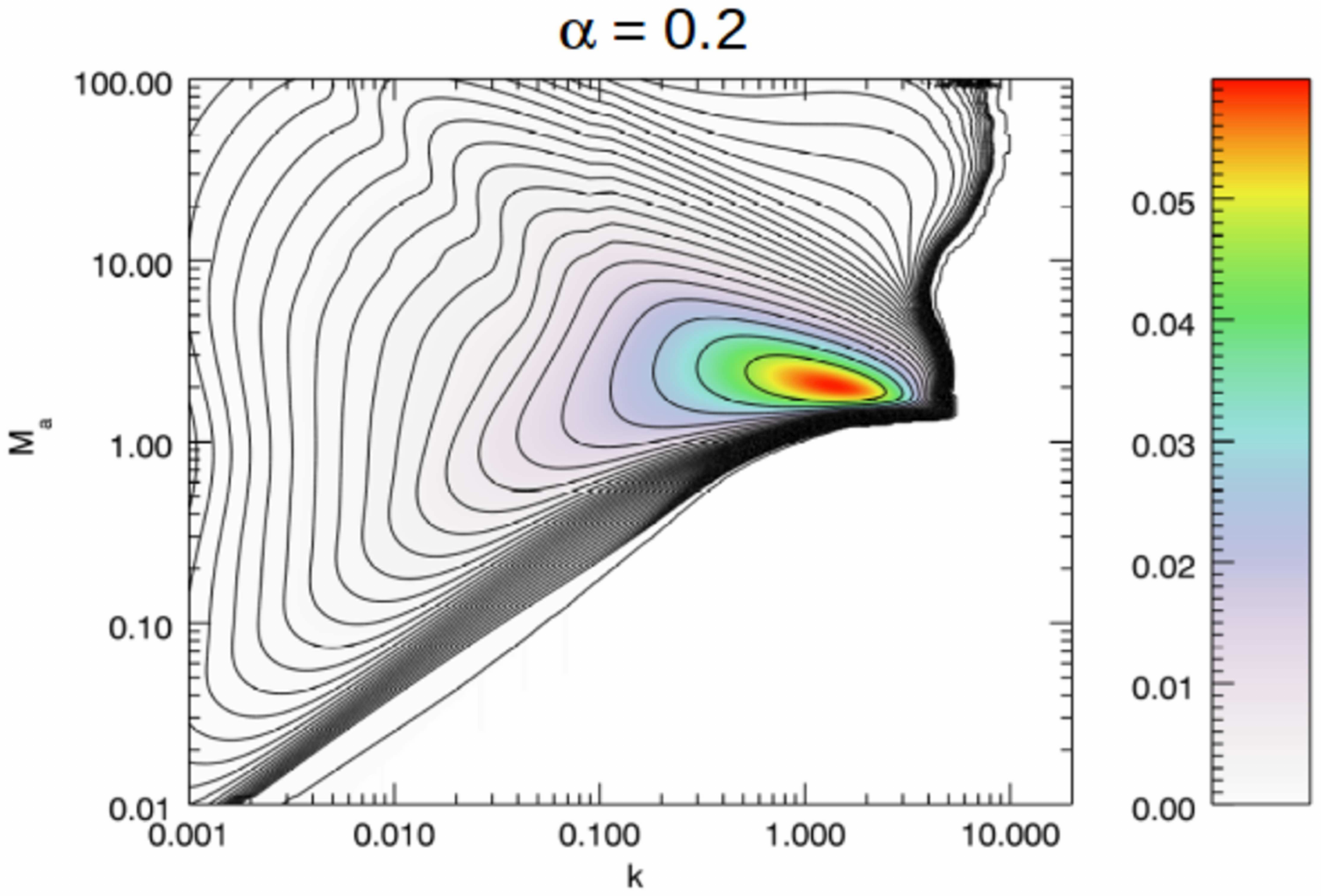}
\caption{Growth-rate contours of instabilities in a zero pressure, magnetized, relativistic jet as a function of the wavenumber, $k$, and the ratio of magnetic to kinetic energies, $M_a$, for a series of initial jet structures \cite{bo19}. In all the plots, the jets have Lorentz factor 10. The left column shows models with no rotation, whereas the right column shows models with rotation (the parameter $\alpha$ gives a measure of rotation speed, see \cite{bo19}). The upper panels give the solutions for $P_c=10$, i.e., with a dominating poloidal field (see the text), the middle panels give the solutions for $P_c=1$, and the bottom panels for $P_c=0.1$, i.e., dominating toroidal field. The region with $M_a> 1$ displays the KHI solutions, and the CDI appears at low values of this parameter, i.e., for magnetically dominated jets, and mainly when the toroidal component of the field is more important (courtesy of Gianluigi Bodo).}\label{fig4}
\end{flushleft}
\end{figure}

Interestingly, the wavenumbers larger than $m/P_c$ ($P_c$ is the value of $P$ at the axis, which converts to $\gamma\,P_c$ in the bulk plasma reference frame) are stabilized, and only those values that are smaller show unstable solutions (see Fig.~\ref{fig4}), for the magnetic field distributions considered. As a consequence, the authors find that a purely longitudinal field is only KH unstable (the KHI growth rates do not depend strongly on the pitch parameter).

 Another relevant parameter is the ratio of total gas to magnetic energy densities, $M_a$ (which in the case of cold jets, as considered by Bodo and collaborators, reduces to the kinetic to magnetic energy density ratio): for large values of this parameter, i.e., matter dominated flows, the jets are only KH unstable; the jet is stable with respect to KHI below a certain value of $M_a$ that depends on the pitch parameter, with this threshold moving to larger values of $M_a$ for larger values of $P_c$ (i.e., smaller toroidal fields). For dominating toroidal fields, this threshold moves to smaller values of $M_a$ and the KHI and CDI curves merge in the solution plane (see Fig.~\ref{fig4}). As a global conclusion, one can state that a strong poloidal field stabilizes the jets, unless $M_a$ is large, i.e., unless the plasma is highly super-Alfv\'enic \cite{bo13}. Irrespectively of the poloidal field intensity, the becomes only KH unstable in the case it dominates the field structure. We have to note, nevertheless, that a strong poloidal field would set demanding limits on the values of the field at injection ($B_z\propto R_j^{-2}$). 
 
 Unlike for the KHI, the CDI growth rates are basically independent of $M_a$, and increase with $B^\phi$. The resonant condition ($\mathbf{k \cdot B} = 0$) makes the CDI growth rates change according to the value of the pitch parameter, which can vary across the jet, for instance, when the field distribution results in a return current at the shear layer \cite{bo13}. This situation results in a non-straightforward relation between the growth rates of instabilities with different wavelengths at different radii, with the modes being stabilized where the pitch parameter changes fast, i.e., in the region where the current is located, but the modes can still grow inside and outside of this region. This is easily explained using Eq.~\ref{eq:res}: if $B^\phi \rightarrow 0$ when $r \rightarrow 0$, the resonance can only be obtained for small wavenumbers, whereas the opposite is true for large values of $B^\phi$ (when $r\rightarrow R_j$). 

     
       Kim et al. \cite{kim17,kim18} have studied the stability properties of jets without current-sheets at the jet boundary. The magnetic field distribution that these authors use is based on a {\em tractable} equilibrium solution for a current-sheet free jet, extracted from the set of solutions derived in \cite{gou12}, with both the axial and toroidal components becoming zero at the jet boundary. The distributions also imply zero net poloidal current along the jet cross section. Their approach allows for non-zero pressure plasmas, making it more general. The authors discuss the stability properties of such jets in the linear regime, confirming the stabilizing role of the Lorentz factor, in terms of decreased growth rates at all wavelengths, in agreement with the RHD case \cite{pe04a,pe05}. With the proposed magnetic field configuration, increased magnetization stabilizes the pinching fundamental mode and the helical (or kink) fundamental mode at short wavelengths. However, it does not affect the growth rates at long wavelengths for the helical mode, leaving room for the growth of these disruptive modes. An extension of their work to velocity-sheared jets has shown that the presence of a shear layer causes a drop in the growth-rates of the short-wavelength modes (as it is also the case in hydrodynamical jets), but again it does not show a strong effect on the long-wavelength, fundamental kink modes \cite{kim18}. In \cite{kim15}, the authors claim that the magnetic field configuration given by Bodo et al. \cite{bo13} implies a constant value of the pitch parameter across the jet, which makes the CDI to become relevant according to Eq.~\ref{eq:res} (see also Fig.~\ref{fig4}), and also to be suppressed from a given wavenumber ($k_z \sim 1/P$). In contrast, if the pitch parameter changes across the jet, such a limit in the development of CDI at high wavenumbers disappears, and the CDI modes show up at all wavenumbers, albeit less enhanced than in the distributions used in \cite{bo13,bo16,bo19}.  
       
     Numerical simulations by Mizuno and collaborators for static columns \cite{mi09} have shown that CDI is stabilized by a dominating axial field and that the pitch angle profile can also change the growth of the unstable modes: the growth rates of the kink mode are reduced by an increase of the pitch parameter (defined as in \cite{bo13}) with the radius (i.e., if the toroidal field becomes relatively weaker), and they increase if the pitch parameter falls with the radius. In \cite{mi11,mi14}, the authors found that the location of the velocity shear in a sub-Alfv\'enic jet can also determine the growth rates of CDI modes and whether kinks are very slowly propagating or move downstream at faster speeds, with larger velocity shear radii (with respect to location of the peak value of the toroidal field) reducing the mode growth rates and also making the kinks slower.
   
    The possible development of the pressure-driven or 'Z-pinch' instability in jets has also been pointed out in \cite{beg98} (see also, e.g., \cite{ke00}, for rotating MHD jets, and \cite{lo08}) with pressure and magnetic gradients across curved magnetic lines as those in toroidal/helical configurations. The pressure gradient arises naturally to generate equilibrium configurations when magnetic pressure and tension have to be compensated (see, e.g, \cite{ma15} and references therein). In such a situation, the displacement of a magnetic loop can trigger the growth of the instability if the toroidal field increases with radius, consistently with the conclusions derived for the CDI (see above). This instability can destroy the concentric structure of the magnetic field close to the jet spine, endangering the possible collimating role of the magnetic field. Maximum growth-rates correspond to small axial wavenumbers in this instability, so no global (large-scale) effects would be visible in the jet structure, other than the probable radiative outcome of dissipation. Nevertheless, it was pointed out \cite{lo08} that a velocity gradient (in the region where the instability develops) can be a compensating mechanism against the pressure-driven instability. 
                  
\paragraph{\emph Stability of magnetized, rotating jets}                      
     
  The introduction of a rotation velocity in the flow represents a stabilization effect for CDI, mainly for more magnetized jets \cite{bo19}, but does not strongly affect the KH unstable modes. For small values of $P_c$ (i.e., dominating toroidal field), the CDI instability appears at smaller wavenumbers (in other words, larger wavenumber are stabilized) with decreasing $M_a$, which means that rotating, strongly magnetized jets are more stable against the CDI (see also \cite{ip94,ip96,lyu99,to01}).\footnote{Istomin \& Pariev \cite{ip94,ip96} studied the development of the CDI in a force-free jet. The authors found that the shear in the toroidal field plays a strong stabilizing role both for axisymmetric and non-axisymmetric perturbations, in the absence of gas pressure, by prohibiting radial convective motions.}
  
  In the case of rotating flows with helical fields, an equivalent to Parker instability arises in the jets, where the role of gravity in that case is played by the centrifugal force (see \cite{bo16,bo19} and references therein). The growth rates of this instability increase with the square of rotation frequency, $\Omega$, and are independent of the wavenumber. The instability shows up at low wavenumbers for the toroidal buoyancy mode and at large wavenumbers for the poloidal one. In \cite{bo16,bo19}, the authors conclude that the modes are stabilized by large values of $M_a$ (i.e., particle dominated) and also in the magnetically dominated limit $M_a \rightarrow 0$ (i.e., they only appear at intermediate values of $M_a \sim 1-10$) for the studied cases. Millas et al. \cite{mil17} revisited the simulations of rotating cylindrical flows run in \cite{mk07,mk09} and showed that the inclusion of a toroidal component stabilized the development of RTI (or CFI) in rotating jets.

\begin{table}
\centering
\tablesize{\normalsize}
\begin{tabular}{lccccccc}
\toprule
  & & $R_j$ & $\rho_j \gamma_j$ & $B^z$ & $B^\phi$ & $\delta_s$ & $v^\phi$ \\
  \midrule
 KHI & &  $\checkmark$ & $\checkmark$ & $\checkmark$ & $-$ & $\checkmark$ & $\checkmark$ \\
CDI & &  $\checkmark$ & $\checkmark$ & $\checkmark$ & $\times$ & $\checkmark$ & $\checkmark$ \\
RTI/CFI & & $\times$ & $\times$ & $-$ & $\checkmark$ & $-$ & $\times$ \\
\bottomrule
\end{tabular}
\caption{Summary of the stabilizing/destabilizing factors for each of the most relevant/studied instabilities in relativistic jets so far. A \checkmark symbol indicates a stabilizing effect and $\times$ indicates a destabilizing effect as the parameter grows. From left to right, the parameters are: jet radius (expansion), jet inertia, poloidal field, toroidal field, shear-layer width, and toroidal velocity. The $-$ sign indicates that there is no effect or that it is ambiguous, depending on factors like radial distribution. The table only pretends to give general and orientative trends but different radial distributions or combinations can change the trends given (see the text).} \label{tab:1} 
\end{table}
 
\subsection{The non-linear regime (or 'Then, why are (some) jets so stable?')} \label{sec:nlr}
  
   In the previous section, I have described the instabilities that grow on the configurations that have been studied by different authors, and the conditions under which the unstable modes can have larger or smaller growth-rates. In summary, from the studies performed so far we can state that RMHD jets are more stable when 1) the toroidal component of the field is smaller, 2) the pitch parameter changes within the jet, 3) the jet rotates with relatively small rotation frequencies, 4) in those parts of the jet cross-section where the toroidal field falls with the radius, 5) the jets are surrounded by shear-layers and/or expand, and 6) have more inertia. Table~\ref{tab:1} summarizes these general trends for the KHI, CDI and RTI/CFI.

  However, from the description given, one can extract the (correct) impression that there is no completely stable configuration and that jets are prone to destabilization under basically any initial distribution of the physical parameters, the difference being the time/distance in which the unstable modes grow to nonlinear amplitudes. And then conclude that there is something that we do not understand in jet physics because, according to linear theory, all jets should be destroyed by instabilities at some point. 

     The incautious reader has to be thus warned about the limits of linear theory when applied to extragalactic jets. The linear studies actually face the difficulty of having to impose very specific initial configurations that equilibrate the transversal profiles of jets under several {\em a priori} assumptions about the magnetic field distribution across the jet. This situation is forced by the impossibility to get a clear picture of the magnetic field structure in jets from observations, due to relativistic and projection effects \cite{ly05,fue18}. In addition, linear analysis is usually based on the consideration of infinite jets. We have seen, within this assumption, that even if an unstable mode has a large growth rate under certain jet parameter configuration, this may rapidly change under changing conditions (see the previous section), which might be the case either across an inhomogeneous jet (for short wavelength modes), or along a jet with changing properties in distance (e.g., due to expansion, mild entrainment..., see Sect.~\ref{sec:int}). 
      
    Moreover, there is a possibly generalized misconception caused by the extension of growth rates in the linear regime to their destabilizing role: we tend to think that the most disruptive instability modes are those showing larger growth rates and that the most unstable jets (and thus more prone to disruption) are those for which the growth rates are larger. However, the link between the linear and the non-linear regime has been shown not to be straightforward by numerical simulations \cite{pe05,pe10}, and even fast growing modes may be non-disruptive (e.g., see Fig.~\ref{fig1}, \cite{pe07}, and see also \cite{kad08} for an application to the radio galaxy 3C~111). Such fast-growing, non-disruptive modes are necessarily short wavelength instabilities that induce small scale changes in the flow shear-layer. As a result, the jet structure and properties (velocity, composition, magnetization...) may change due to dissipation and minor entrainment at the jet boundary, but the outcome would still be considered a collimated, long-term-stable jet.

     In this respect, the pressure-driven instability \cite{beg98} could disrupt the toroidal structure of the flow, possibly dissipating magnetic and kinetic energy, but if the jet retains (at least to some extent) the axial field structure, this component can reduce the growth rates of the KHI and completely stabilize both the CDI.\footnote{Actually, FRII jets show polarization structure at kpc scales that is compatible with a dominating axial component \cite{bri84,bri94,beu18}, thought to be produced by shearing.} However, it is implausible that the poloidal field dominates the jet dynamics at large scales on the basis of magnetic flux conservation (see above). 

     Not all short wavelength instabilities are harmless to jet collimation. Actually, small-scale RTI or CFI modes have been proposed as the cause of the progressive deceleration in FRI jets \cite{ma17,gk18a}. In this case, the difference lies on the jet properties: both the RTI and CFI require jet expansion with relatively large opening angles, i.e., slower, overpressured (hot) jets propagating in dilute media or through steep pressure gradients.   

   Longer wavelength instabilities do develop in jets and we can explain the structures forced by these growing large-scale perturbations, considering global jet properties and the small jet opening angles (actually, jets can freely expansion and still have a small opening angle, $\sim 1/\gamma$, to fulfill the linear theory basic assumptions). This idea has been successfully applied to a number of well-known radio galaxies and quasar jets (e.g., \cite{lz01,har87,har03,har05,har11,pe12,vg19}), typically for cases in which the observed wavelengths are long enough, $\lambda \gg R_j$, not to depend critically on the jet internal fine structure. It has also been shown, by means of numerical simulations, that the instabilities can develop and show up only at given regions of the jet cross-section and be detected only by observations at the frequencies that reveal those regions \cite{pl06}. 
   
   Regarding the long-term jet stability, I review and enlarge here the list of arguments that can make jets stable for long distances. First of all, I would like to clarify the meaning of 'jet disruption' or 'long-term stability' in order to set the frame of the discussion. From this author's point of view, a jet is disrupted when it is strongly mass-entrained and decelerated, losing collimation and acquiring plumed structure in radio images. Under this point of view, jets can develop instabilities reaching the non-linear regime, and even be entrained and decelerated, without becoming a 'disrupted jet'. In other words, a jet is not disrupted unless it is strongly decelerated and losses the inertia that contributes to its collimation, even if its composition, transversal structure or magnetization change. Therefore, we have to understand 'long-term stability' as the ability of jets to reach hundreds of kiloparsecs without being decollimated. We can think, for instance, about the prototypical jets in Cygnus A and consider them as long-term stable FRII jets. However, its jets show an obvious large-scale kink close to the hot-spots, so 'stability' has to be taken as a relative term. 
      
   A very relevant, and stabilizing, difference of the growth of instabilities in relativistic flows was pointed out by M. Hanasz \cite{ha97}, who predicted that the amplitude of the KH unstable wave would cease to grow as soon as the amplitude of the velocity perturbation approaches the speed of light in the jet reference frame. This was later confirmed by numerical simulations \cite{pe04a,pe04b,pe05}. The physical limit imposed by the speed of light shows to be efficient in avoiding the indefinite growth in amplitude of the unstable modes, thus providing the flow with a chance to avoid disruption. Nevertheless, the simulations also showed that for a large area of the parameter space (relativistic Mach number vs Lorentz factor, see \cite{pe05}) the instabilities reach an amplitude that is enough to disrupt the flow before the limit is reached.
   
   Jet expansion can also represent a stabilizing mechanism \cite{har86}: the KHI modes grow when reflected at the discontinuity/shear-layer between the jet and the ambient medium (with the jet acting as a resonant waveguide, \cite{pc85}), so the increased distance to be covered by the wave to reach that boundary increases the growth-length of the mode. In \cite{kim17,kim18}, for instance, the authors compare the growth time-scale of the unstable modes $\tau = 1/|\omega_i|$ with the propagation time-scale $T=K R_j / v_g$, where $K$ is a constant indicating the number of jet radii considered (for an instability to develop to the nonlinear regime), and $v_g$ is the group velocity. The comparison between $\tau$ and $T$ gives an idea of the disruptive potential of a given mode (e.g., if $\tau < T$, the growth rate of the mode is too slow to destabilize the flow). Thus, we can also say that expansion increases $\tau$, favoring jet stability. Actually, expansion has been even proposed as the main factor to explain jet collimation at large scales, via the loss of causal connectivity across the jet if the ambient pressure, $p_{ext} \propto z^{-\kappa}$, falls with $\kappa \geq 2$ \cite{pk15}. 
   
   Surrounding winds formed from the accretion disk \cite{bp82} or thick shear layers can contribute to reduce the growth of instabilities, as shown by linear analysis (see Sec.~\ref{sec:lin}), and also numerical simulations (e.g., \cite{hh03,mi07,mi14}).

\begin{figure}
\centering
\includegraphics[trim={1cm 6cm 1cm 5cm},clip,width=0.35\textwidth]{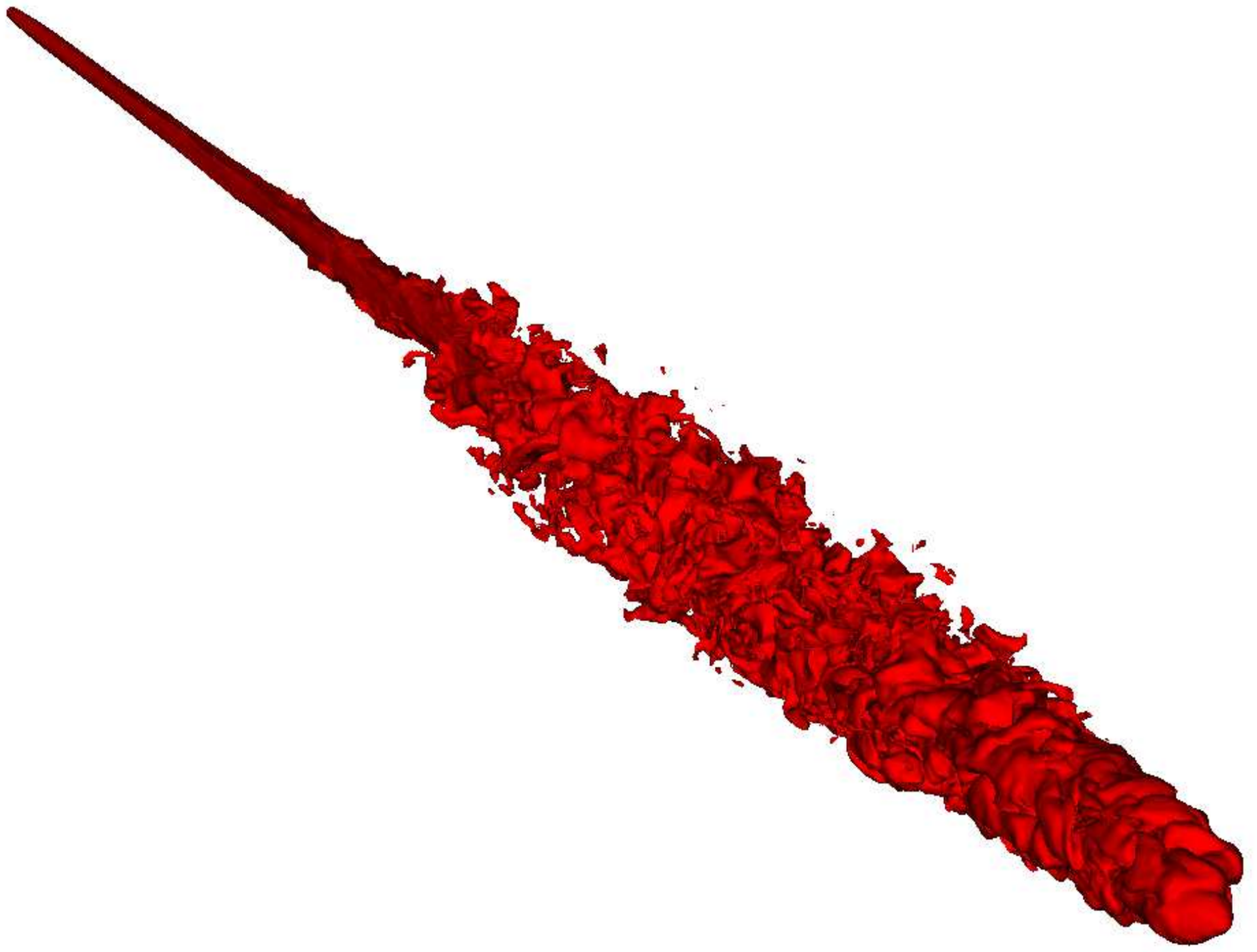} \quad \includegraphics[trim={1cm 13cm 0cm 2cm},clip,width=0.45\textwidth]{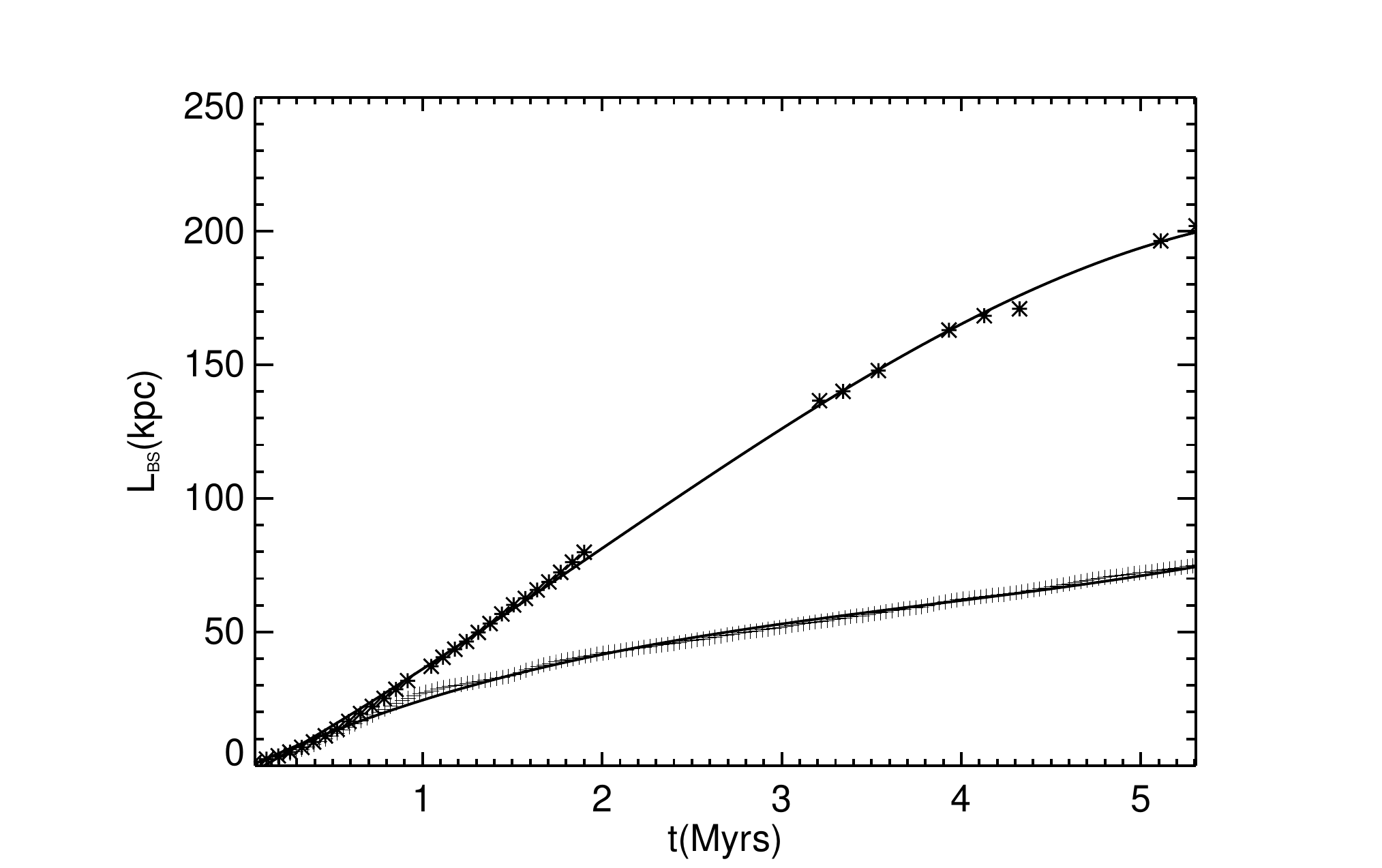}
\caption{(\textbf{a}) The left panel shows the jet-mass fraction 50\% isosurface for a 3D RHD simulation of a powerful relativistic jet propagating through a density/pressure gradient typical of an elliptical galaxy \cite{pe19}. The jet collimation reminds that observed in FRII jets. (\textbf{b}) Jet head position versus time for the 3D simulation shown in the left panel and a 2D axisymmetric simulation with the same injection parameters, for comparison; the lines show polynomial fits to the data. The 3D jet propagates faster due to head acceleration produced by the obliquity of the front shock as compared to the planar shock typically observed in 2D simulations (see \cite{pe19} for a detailed discussion).}\label{fig5}
\end{figure}

     Finally, there is a key point in this discussion: jets gain stability with respect to the KHI, for instance, as they evolve through an ambient medium that becomes more dilute (this also applies to the cocoon, with a decreasing density as it expands, see, e.g., \cite{har86,pk15}). Therefore, those jets with larger head advance speeds can reach large distances before any instability can trigger their disruption, which has to be considered in parallel to the fact that larger bulk Lorentz factors result in smaller instability growth-rates. Furthermore, a recent paper \cite{pe19} shows that small-amplitude oscillations of the jet head caused by linear helical unstable modes can cause accelerated head velocity down density/pressure gradients across and out of active galaxies (see Fig.~\ref{fig5}). Therefore, the old Catalan saying {\em qui dia passa, any empeny}, with approximate translation 'that who survives a day, pushes a whole year', can be applied to long-term-stable jets. In other words, jets are unstable systems but, in some cases, manage to reach long distances before any growing instability endangers their collimation (e.g., FRII sources), whereas, in others have enough inertia to develop to large scales (through a decreasing density ambient medium) even if mass-loaded and decelerated, i.e., 'disrupted'. In this case, the jets expand in the turbulent regime, but can still propagate to large distances in external density/pressure gradients (\cite{bi84}, e.g., large-scale FRI sources as the radio jets in 3C~31, which reach $\geq 200~kpc$). 
 
\section{Jet star/cloud interactions}\label{sec:int}

   On top of instabilities, there is another efficient mechanism to convert magnetic or kinetic energy into internal energy and particle acceleration,\footnote{The direct conversion of magnetic energy into internal energy implies dissipative mechanisms beyond an ideal MHD description.} which is direct entrainment. It can be caused by the growth of small-scale instabilities causing the formation of turbulent layers that propagate toward the jet center, or by the direct impact of the jet with a stellar wind or a gas cloud. In the latter, the supersonic nature of jets triggers shocks and turbulent mixing tails of shocked jet and wind plasma (see the left panel in Fig.~\ref{fig7}). This scenario can thus also be a source of particle acceleration via first-order Fermi acceleration in shocks, Fermi II acceleration in the turbulent mixing tails, or shear acceleration (e.g., \cite{rl18} and references therein).


\begin{figure}
\centering
\includegraphics[trim={0cm 8cm 0cm 8cm},clip,width=0.8\textwidth]{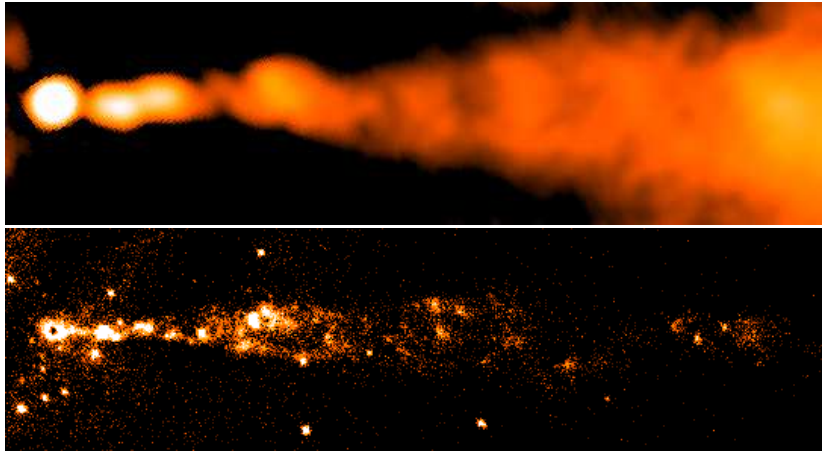} 
\caption{Centaurus~A jet as observed by the VLA at 5~GHz and by Chandra at X-rays. Adapted from \cite{hc11} (courtesy of Martin Hardcastle). The X-ray knottiness of this jet has been interpreted as possible jet-massive star interactions. In the radio, the emission shows a diffuse structure.} \label{fig6}
\end{figure}   

   It was first noted in \cite{bp97} that the interaction between a star and the jet flow could be responsible for gamma-ray production in extragalactic jets. Before, these interactions had been claimed to be possibly responsible for the knotty structure of M87 \cite{bk79}. Although this is not the most accepted scenario for M87, the interaction has been proposed to explain gaps in radio emission at the sub-parsec scale in Centaurus~A \cite{mu14} and its knottiness at kiloparsec scales \cite{wo08,go10,hc11} (see Fig.~\ref{fig6}), and also as an explanation for the gamma-ray emission (\cite{bba10,ara13,wyk13,br15,wyk15,vi17,tab19}) and flares recurrently observed in radio galaxies (M87, \cite{aha06,al08,ac09,ali12,abr12,bba12}), or blazars (3C~454.3, \cite{kh13}, or CTA~102, \cite{za17}). 
   
   Numerical simulations of this scenario, in the context of relativistic flows, have only been performed in the RHD limit \cite{br12,dlc16,pe17b}. These simulations have been focused on the dynamical and energetic impact of the interaction, plus the expected radiative output. In \cite{br12}, the authors studied the evolution of jet/star interaction where the star is surrounded by an envelope or wind before the system is stabilized (see below). The simulations showed that these envelopes can be completely disrupted before the star crosses the jet (see also \cite{pbr12} for relativistic flow-cloud interactions in X-ray binaries), thus incorporating a large amount of hadronic material in the flow. The shock is stabilized at a distance from the star that is given by the equilibrium between the jet and the stellar wind ram pressures \cite{ko94,dlc16}:
 
\begin{equation}\label{eq:rs}
R_s=\sqrt{\frac{\dot{M}_w\,v_w} {4 \pi\,\rho_j\,\gamma_j^2\,v_j^2 }},
\end{equation}  
 where $R_s$ provides the location of the contact discontinuity, $\dot{M}_w$ is the stellar wind mass flux, and $v_w$ is the wind velocity, and the subscript $j$ refers to jet values (note that the velocity is in physical units in this expression). During the phase previous to equilibrium, the shocked stellar wind forms a bow-shock structure and a cometary tail along which the shocked wind is accelerated (see Fig.~\ref{fig7}). If the tail is destabilized, the shocked jet and shocked wind flows can eventually mix. As a result of the jet-star/cloud interactions, the resulting jet transversal structure probably becomes significantly irregular, with slower/colder streams of gas continuously flowing and mixing with the jet flow and thus becoming an excellent location for particle acceleration. The key point is to know in which scales and amounts we expect this to happen. On the one hand, the high mass and inertia of the cometary tail provide it with remarkable stability \cite{br12,pe17b}, but 3D simulations show that the asymmetric perturbations forced by the propagation of the star through the jet probably induce non-linear oscillations that can already destroy the tail close to the interaction region \cite{pe17b} (see the right panel in Fig.~\ref{fig7}). Therefore, turbulent mixing is expected to take place not far downstream from the interaction site.
  
\begin{figure}
\centering
\includegraphics[trim=6.5cm 8.cm 11.5cm 9.5cm,width=0.1\textwidth]{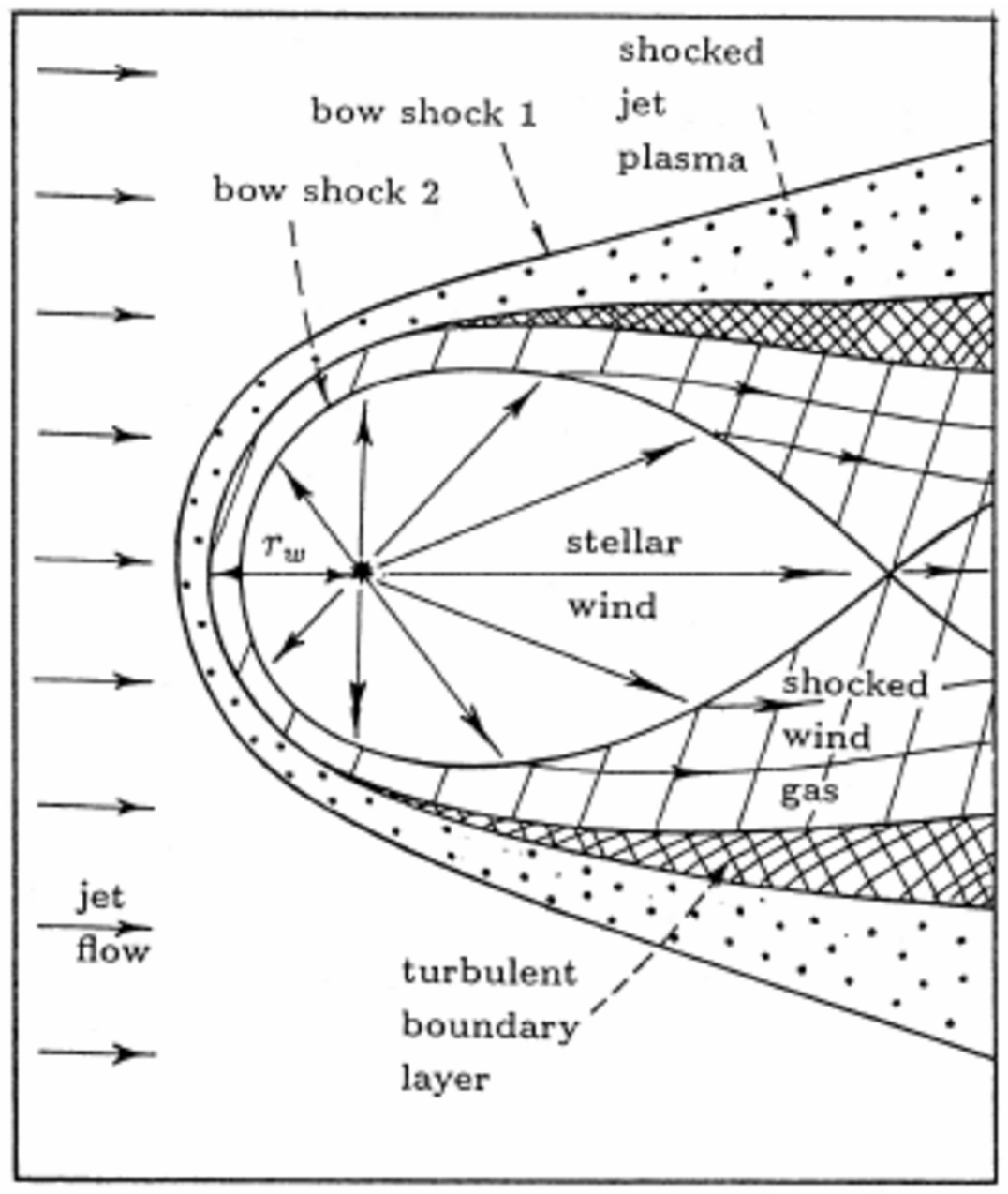} \hspace{4.cm} \includegraphics[trim={0cm 2cm 0cm 4cm},clip,width=0.45\textwidth]{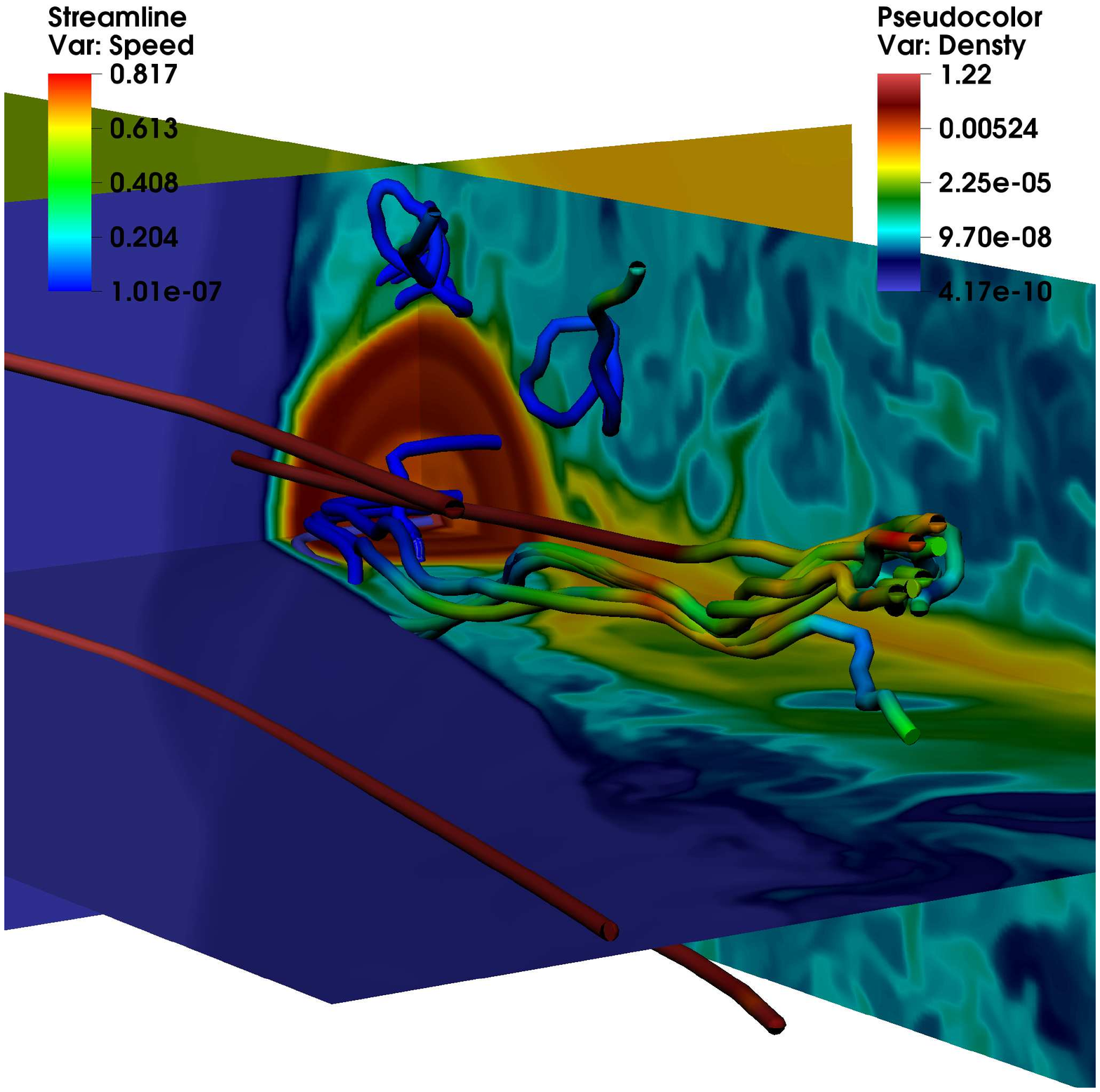} 
\caption{(\textbf{a}) Schematic plot of jet-stellar wind interaction \cite{ko94}. (\textbf{b}) Rest-mass density distribution from a 3D RHD simulation of a star/wind bubble system entering into a jet \cite{pe17b}. The image shows three cuts across the grid, displaying the position of the wind injector (star). The color scales indicate the speed (top left) and the rest-mass density (top right). The star enters the jet from the right to the left, which is occupied by the jet flow. The jet flow appears in blue color, with a velocity directed towards the bottom of the image. The current lines show, on the one hand, the deviation and deceleration of the jet flow as it crosses the bow-shock, and, on the other, the shocked wind material being dragged downstream and forming an unstable tail.}\label{fig7}
\end{figure}


\begin{figure}
\centering
\includegraphics[width=0.45\textwidth]{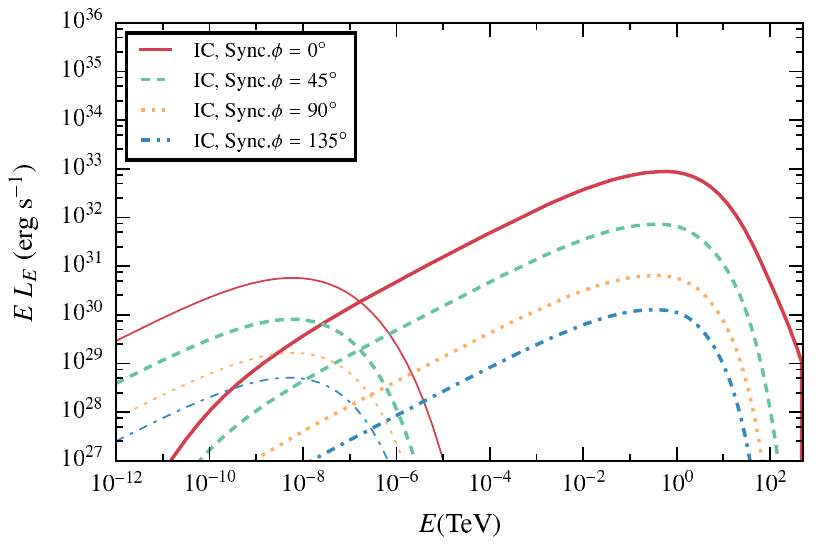} \quad \includegraphics[width=0.45\textwidth]{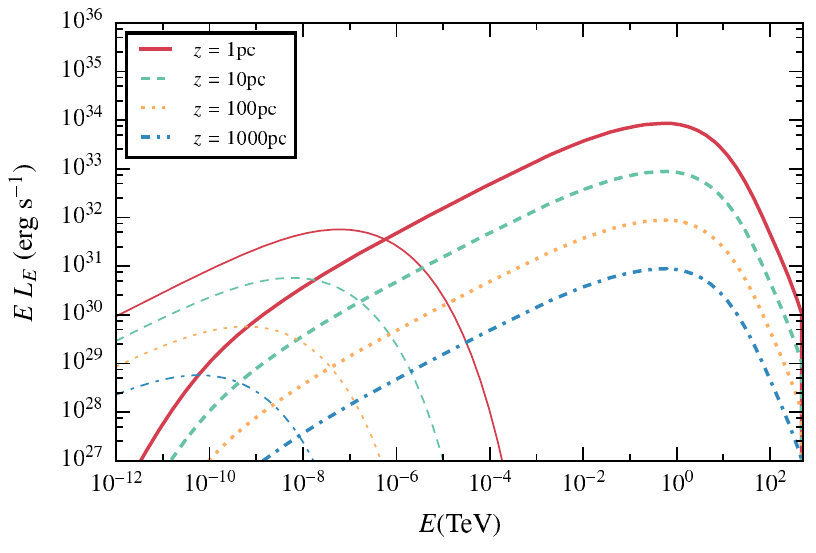}
\caption{(\textbf{a}) Resulting spectral energy distributions (SED), including synchrotron and IC radiation, of the equilibrium state in a jet-star (red giant, $\dot{M}_w = 10^{-9}\,{\rm M_\odot/yr}$) interaction at 10~pc from the central engine for a weakly magnetized jet, as derived from different viewing angles \cite{dlc16}. An increase of the magnetization increases strongly the synchrotron flux but barely affects the IC radiation. (\textbf{b}) Same as (a), but for a fixed viewing angle of $\phi = 0^\circ$ and different interaction locations along the jet. The farther from the central engine, the weaker the resulting emission.} \label{fig8}
\end{figure}

  A very detailed entrainment/deceleration model, based on a previous study of the stellar population in Centaurus~A was developed by Wykes et al. \cite{wyk15}, who estimated that there could be $10^8-10^9$ stars within the jet, entraining a total of $\sim 10^{-3}\,{\rm M_\odot/yr}$, and could reproduce the spectral distribution of the kpc-scale jet from radio to X-rays with this model by means of the synchrotron emission produced by the interactions (via the Fermi I mechanism). The radiative output of the interaction depends on the location along the jet, with close impacts to the SMBH possibly explaining gamma-ray flares via proton-proton interactions, as proposed for M87 \cite{bba12}, but even with collective downstream interactions underpredicting the steady gamma-ray flux detected from this galaxy \cite{vi17}. Vieyro et al. \cite{vi17} studied the collective synchrotron plus IC emission caused by jet-star interactions (with the seed IC photons considered to be infrared photons in the case of starburst galaxies and starlight plus CMB for galaxies with older stellar populations, e.g., M87). The authors applied their model to different cases, including the FRI radiogalaxy M87, and concluded that the diffuse radiation emitted by these interactions can be strong at X-rays via synchrotron, although the X-ray emission is more structured in the case of M87 (cf. other FRI jets showing diffuse emission \cite{kh12,lb14}). Furthermore, the gamma-ray emission produced via IC radiation could significantly contribute to the persistent gamma-ray flux received from the studied galaxies. According to Torres-Alb\`a \& Bosch-Ramon \cite{tab19}, the dependence of the radiative output with the jet power is relatively weak in the studied range ($10^{43}\,{\rm erg/s}<L_j<10^{45}\,{\rm erg/s}$), and synchrotron photons up to 100~MeV could be produced for  a weak magnetic field (with respect to equipartition) or by bright events added to the persistent background in the case of stronger field intensities. The authors derived total emitted synchrotron luminosities $\gtrsim 10^{40}-10^{42}\,{\rm erg/s}$ and inverse Compton $\sim 10^{39}-10^{40}\,{\rm erg/s}$ as a result of the collective interaction of jets with a red-giant population. The role of Doppler boosting on the received flux was also highlighted in these works (see \cite{br15} for the case of blazars), with the obvious result that misaligned sources are more difficult to detect. A more exotic scenario is the interaction of the jet with a supernova explosion \cite{vie19}, which could result in stronger gamma-ray emission (IC of infrared photons or synchrotron self-Compton producing up to $10\%$ of the jet power for jets in the range $10^{43}-10^{44}\,{\rm erg/s}$) collimated along the direction of the flow by Doppler boosting, if the jet-supernova interaction occurs at tens of parsecs (the region where a larger number of explosions could be expected) and the ejecta is accelerated downstream of the jet to a high Lorentz factor. This contribution could be steady in the case of galaxies with high star formation rates and weak blazar jets. The model considers that the shocked supernova remnant does not mix with the flow along a sufficient distance, so it obviates the jet deceleration, which, according to the authors, would eventually occur farther downstream from the emitting region ($\sim$~kpc). 
  
   Figure~\ref{fig8} shows different spectral energy distributions for a jet-red giant interaction at the equilibrium stage (i.e., once all the outer wind layers have been eroded and the interaction is located at $R_s$) as observed from different viewing angles (left panel) and at different locations along the jet (right panel, \cite{dlc16}). The radiative output is modest, but it has to be kept in mind that it is the collective emission of all jet-star interactions that can be observed by our detectors. Altogether, the different works coincide on the efficient local conversion of kinetic energy into internal energy at the interaction region between the jet and red giants or massive stars and the plausibility of GeV - TeV radiation generated from these regions. This scenario thus represents a possible explanation, along with (i.e., not excluding) the spine-sheath \cite{ta16} or the jet-in-jet \cite{gi09} models, for VHE emission from AGN jets. Finally, the interaction scenario has been also recently proposed as a plausible option to explain the production of neutrinos in blazars via the hadronic processes following the entrainment of protons from stellar winds \cite{kad16,ice18,kei18,mu18,sa18}.


\section{Influence of entrainment on jet evolution}

   In the previous sections I have focused on our current knowledge about the instabilities that can develop in relativistic jets and jet-obstacle interactions, as possible frames in which VHE photons could be produced. In this section, I summarize the global effect that either shear-layer entrainment or stellar mass-load (or both) can have on jet dynamics and propagation.
   
  In a series of papers, Bicknell \cite{bi84,bi86a,bi86b,bi94,bi95}, De Young \cite{dy86,dy93} and Komissarov \cite{ko88,ko90a,ko90b} explained the transition from supersonic to transonic flows and the radiative properties of FRI jets at kiloparsec scales by means of turbulent mixing at the jet shear layer, although using different approaches and assumptions. De Young modeled entrainment and related the deceleration of jets to the properties of the host-galaxy ambient medium, using conservation equations \cite{dy93}. Komissarov modeled the evolution of the post-deceleration (from relativistic to subrelativistic) flow in FRIs as a turbulent, low Mach number flow with thermal and relativistic populations coexisting and evolving (two-fluid model, \cite{ko90a,ko90b}). The model allows for a conversion of thermal into relativistic particles via the coupling provided by turbulent dissipation. Bicknell \cite{bi84} considered that the entrainment into a jet dominated by relativistic particles would convert it into a transonic, turbulent flow and applied this idea to derive the evolution of the electron distribution and magnetic field, to ultimately calculate the evolution in brightness of FRI radio-sources (assuming flux-freezing and adiabatic losses). The concept behind this model is the dissipation of kinetic energy via weak shocks and particle acceleration, modeled by a single losses term in the kinetic energy, plus a gain term in the heat equation. Later, the same author used conservation laws plus the source terms required to account for mass entrainment to model jet deceleration \cite{bi94}. In the same way, Laing \& Bridle \cite{lb02a} explained the jet deceleration via entrainment. The models developed by \cite{bi94} and \cite{lb02a} were later extended by Wang et al. \cite{wa09} to study the development of the turbulent/shear layer towards the jet axis. In summary, the idea of entrainment, deceleration and dissipation caused by turbulence lays behind the evolution of FRI jets. 

    We have seen in Section~\ref{sec:st} that this entrainment could be caused by small-scale KH, RT or CF instabilities developing at the contact discontinuity between the jet and its environment (either a surrounding, slower and denser wind, or the backflow/cocoon generated by jet propagation). However, Komissarov \cite{ko94} also proposed the possibility of mass entrainment by turbulent mixing at the cometary tails generated by jet-stellar wind interactions. The author also showed that the stellar mass input scenario could be considered as a hydrodynamical problem, given the large size of the expected mixing region as compared to the particle gyroradii. Bowman, Leahy \& Komissarov \cite{bo96} performed steady-state simulations of jet evolution including a source term in the conservation equations to emulate the mass-load by stellar winds, considering an old, low-mass population, typical of giant elliptical galaxies that are the usual hosts of jets, with stellar winds $M_w \sim 10^{-12}\,{\rm M_\odot\,/yr}$. The jets considered were very dilute ($10^{-31}-10^{-33}\,{\rm g/cm^3}$) and hot ($10^{11}-10^{13}\,{\rm K}$). They found that these low power jets ($L_j \leq 10^{42}\,{\rm erg/s}$) could be decelerated by stellar winds alone, and a dual behavior in terms of internal energy/temperature: on the one hand, the hotter models cooled down as they expanded and were mass-loaded, whereas, on the other hand, colder models could gain temperature as they entrained and decelerated. The authors claimed that this was due to the dissipation caused by the mass-load.

   Later, in \cite{lb02a}, the authors estimated that the entrainment needed by jet deceleration in the radio galaxy 3C~31 (a powerful FRI jet, $L_j\sim 10^{44}\,{\rm erg/s}$ could not be caused by stellar winds alone, but that continuous mixing at the shear layer was necessary. In contrast, Hubbard \& Blackman \cite{hb06} showed that a single, large star with a powerful wind, such as a Wolf-Rayet could possibly quench a low power  jet ($10^{42}\,{\rm erg/s}$), as an extreme case, but that a collection of stars with weaker winds could efficiently decelerate the jet. They also derived an estimate of the distance at which the initial jet power is completely exhausted by the work done to push the entrained material, which can be expressed in terms of the stellar density and typical wind mass flux as (Perucho et al., in preparation): 
      
\begin{eqnarray} \label{eq:hb}
l_{\rm d} \simeq \frac{1} {\gamma_{\rm j}} \! \! \left(\frac{L_{\rm j}} {10^{43}\, {\rm erg\,s}^{-1}}\right)
\!\! \left(\frac{\dot{M}} {10^{-11}\, {\rm M_\odot} {\rm yr}^{-1}}\right)^{-1} \left(\frac{n_s} {1\,{\rm pc}^{-3}}\right)^{-1}\,  \!\!
\! \left(\frac{R_{\rm j}} {10\, {\rm pc}}\right)^{-2} \!\! \, 10^2\, {\rm kpc},
\end{eqnarray}
where $\gamma_j$ is the jet Lorentz factor, $L_{\rm j}$ is its kinetic power, $R_j$ is the jet radius, $\dot{M}$ is the mean mass-loss rate of the stellar population in the galaxy, and $n_s$ is the number of stars per unit volume. This last parameter is obviously a decreasing function of distance to the galactic nucleus, so the expression stands as a lower limit of the deceleration distance if $n_s$ is kept at its value in the galactic core. 

\begin{figure}
\centering
\includegraphics[trim={0cm 12cm 0cm 2cm},clip,width=0.45\textwidth]{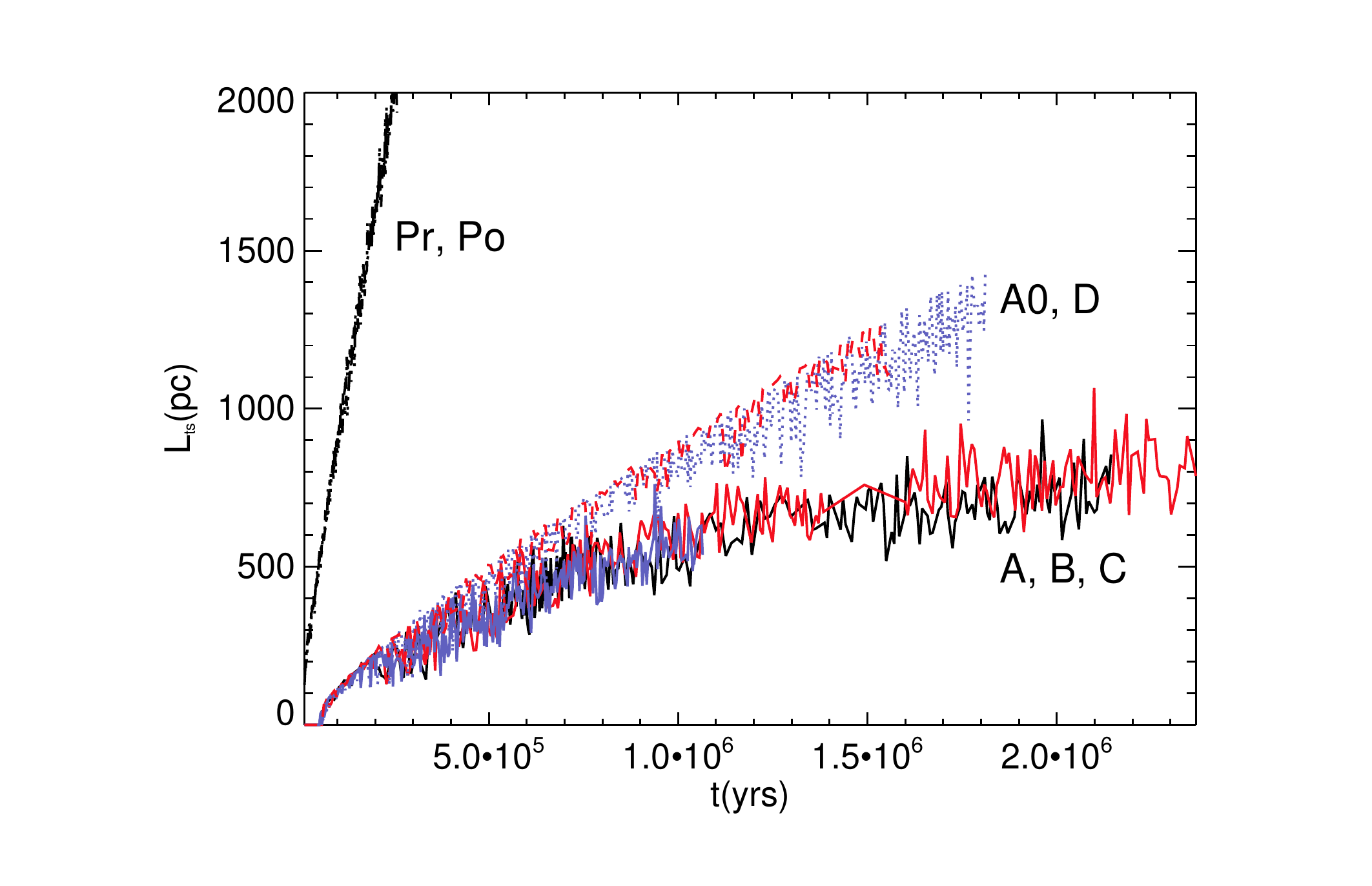} \quad \includegraphics[trim={0cm 12cm 0cm 2cm},clip,width=0.45\textwidth]{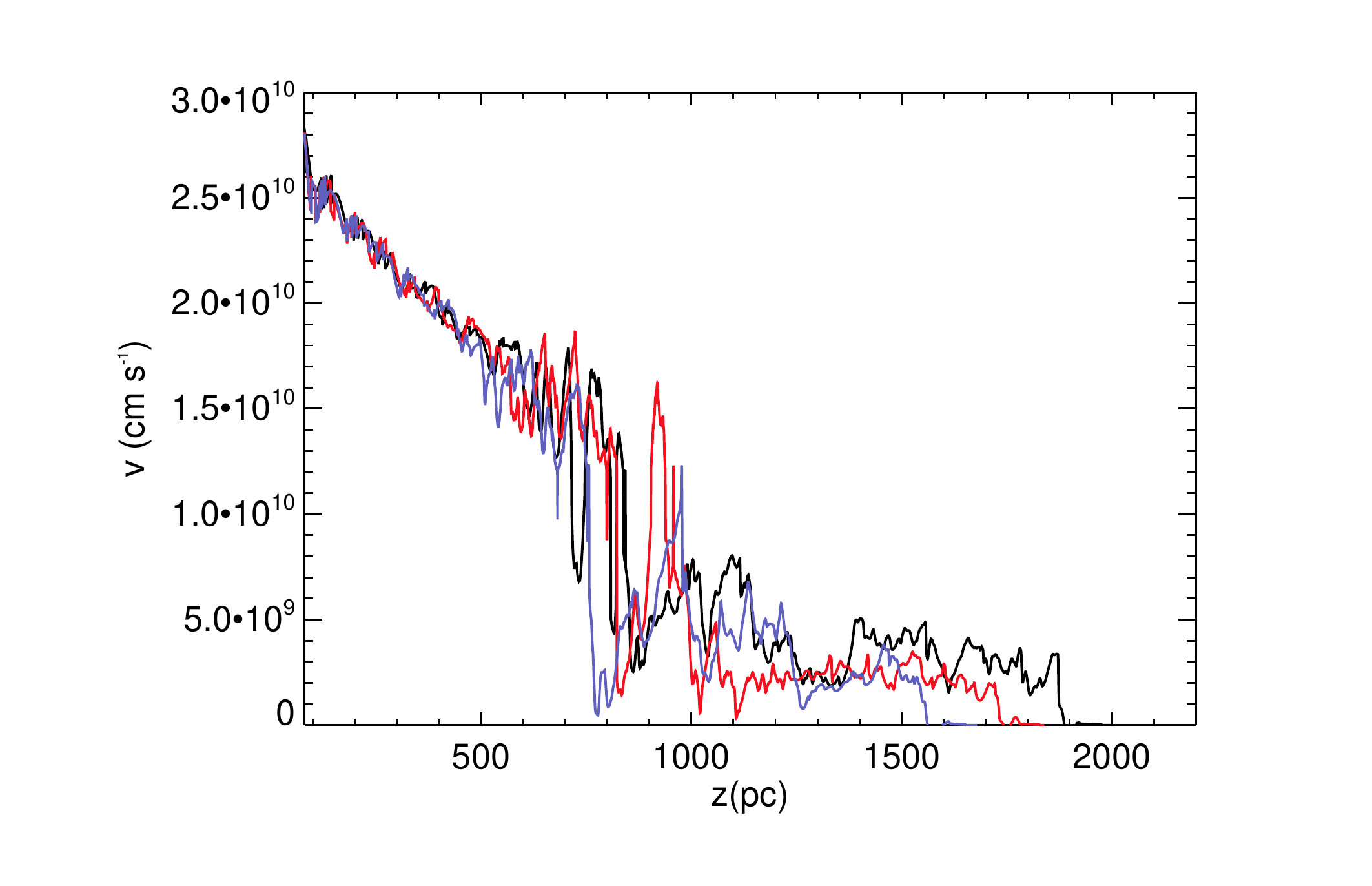}
\caption{(\textbf{a}) Jet head position versus time for different mass-loaded jets. Pr, Po models correspond to jets with power $10^{44}\,{\rm erg/s}$, model A0 is a low power jet ($\sim10^{42}\,{\rm erg/s}$) with no mass-load, used as a fiducial simulation, models A, B, C and D have the same power as A0 but are mass-loaded by stars with $\dot{M}\simeq 10^{-12}\,{\rm M_\odot/yr}$ (models A, B and C) or $\dot{M}\simeq 10^{-13}\,{\rm M_\odot/yr}$ (model D). Models A, B and C only differ in the value of rest-mass density and temperature at injection, with increasing temperature and decreasing density from A to C (ranging from $3\times 10^{11}$ to $3\times 10^{13}\,{\rm K}$ and $3\times 10^{-33}$ to $3\times 10^{-35}\,{\rm g/cm^{3}}$). The expansion velocity is clearly conditioned by entrainment, with one order of magnitude changes in mass-load implying critical changes in jet evolution. (\textbf{b}) Mean jet velocity along the first two kiloparsecs of evolution for models A, B and C (mass-loaded and decelerated). The disruption distance predicted by equation~\ref{eq:hb} for these models is ~200~pc. The difference between the prediction and the simulation is caused by the drop in stellar density with distance. \cite{pmlh14}} \label{fig10}
\end{figure}

   These results anticipated that jet deceleration by stellar mass-load is sensitive to both jet power and the stellar population of the host galaxies. Perucho and collaborators \cite{pmlh14} run RHD simulations of leptonic jets evolving in a galactic ambient medium, using the parameters and set up proposed in \cite{bo96}. The results confirmed the general conclusions derived in the latter work, but also showed that jet disruption by stellar winds is a very sensitive process to the mean stellar wind mass fluxes. In other words, for each jet power there is a minimum mass-load rate required to produce a significant dynamic effect. Fig.\ref{fig10} shows the different effect of mass-load rates differing in a factor 10 ($10^{-11}$ versus $10^{-12}\,{\rm M_\odot/yr}$, \cite{pe14}) on jets with the same kinetic power. The figure also shows the role of jet power, with jets of $L_j = 10^{43}\,{\rm erg/s}$ evolving without remarkable changes with or without mass-load (for the considered stellar wind mass fluxes), as opposed to the low power jets, $L_j \leq 10^{42}\,{\rm erg/s}$. 
   
   It is relevant to remark that, although these simulations introduce the mass-load by means of a source term in the mass conservation equation, the jet/obstacle interactions will induce strong inhomogeneities in jet pressure and velocity across it. The inhomogeneities will necessarily result in changes in emissivity and introduce small-scale instabilities. In \cite{pe17b} and \cite{tab19}, the authors evaluated the mass-load caused by the entrance of a star surrounded by its wind bubble into the jet, concluding that this process could cause enhanced entrainment at the jet shear-layer, if the wind bubble is destroyed before the jet flow reaches equilibrium with the stellar wind at $R_s$ (see Eq.~\ref{eq:rs}). 
   
    Finally, Perucho et al. (in preparation) have run a series of steady-state RMHD simulations of initially leptonic jets, making profit of a one-dimensional approximation to the RMHD equations developed by Komissarov et al. \cite{ko15}, including different mass-load rates, ambient density and stellar number distributions. This work confirms the possibility of efficient jet deceleration, even in the case of typical FRI jet powers of $L_j\sim 10^{43}\,{\rm erg/s}$ depending on the stellar population in the host galaxy ($\dot{M}\geq 10^{-11}\,{\rm M_\odot/yr}$). Again, a dichotomy is found between jets that undergo heating and those that are initially hotter, and cool down as they evolve. Another relevant conclusion of that work is that loaded jets can significantly change their initial composition, and can even become proton dominated, even if not strongly decelerated.     
   
   \begin{figure}
\centering
\includegraphics[trim={0cm 2cm 0cm 4cm},clip,width=0.45\textwidth]{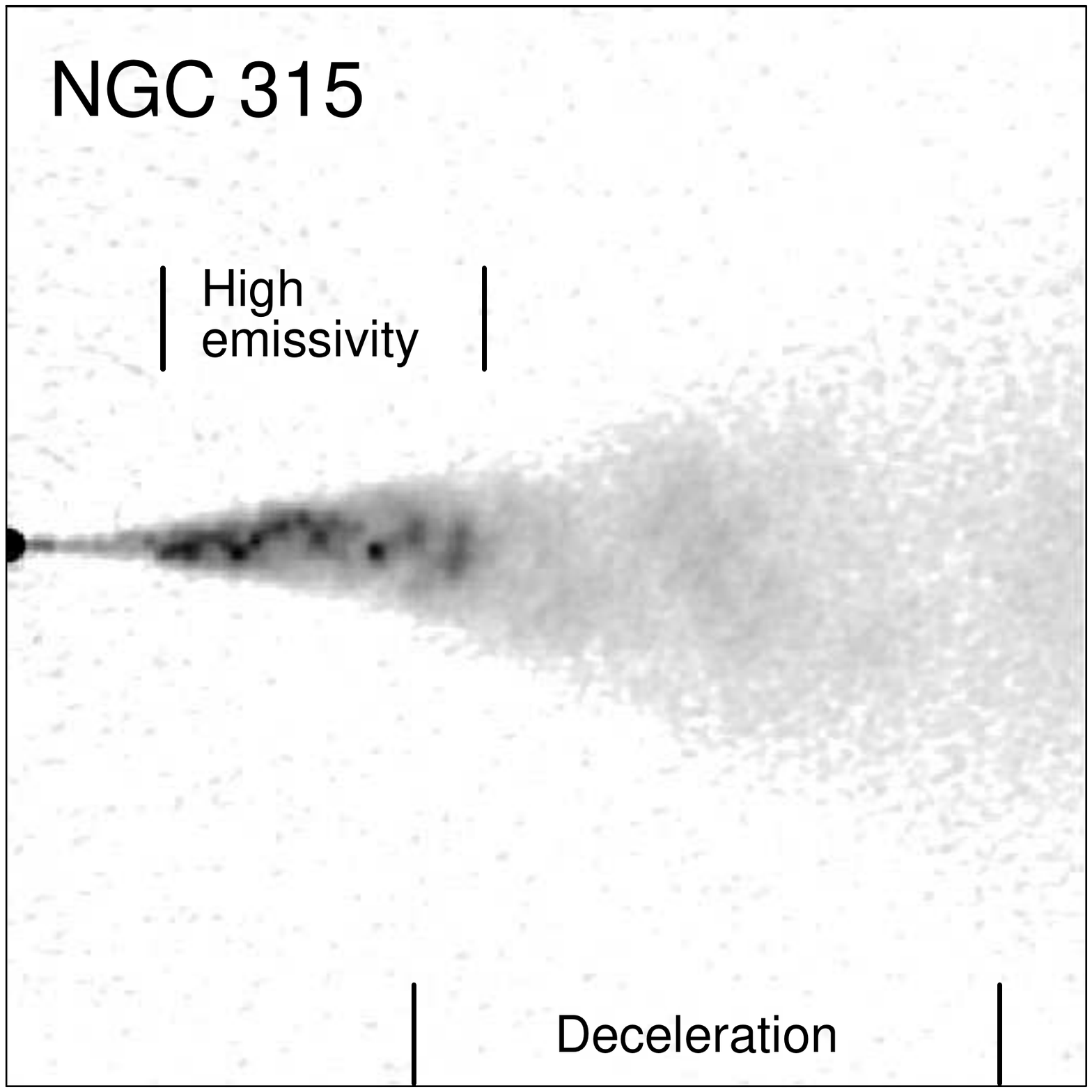}\quad \includegraphics[trim={0cm 2cm 0cm 4cm},clip,width=0.45\textwidth]{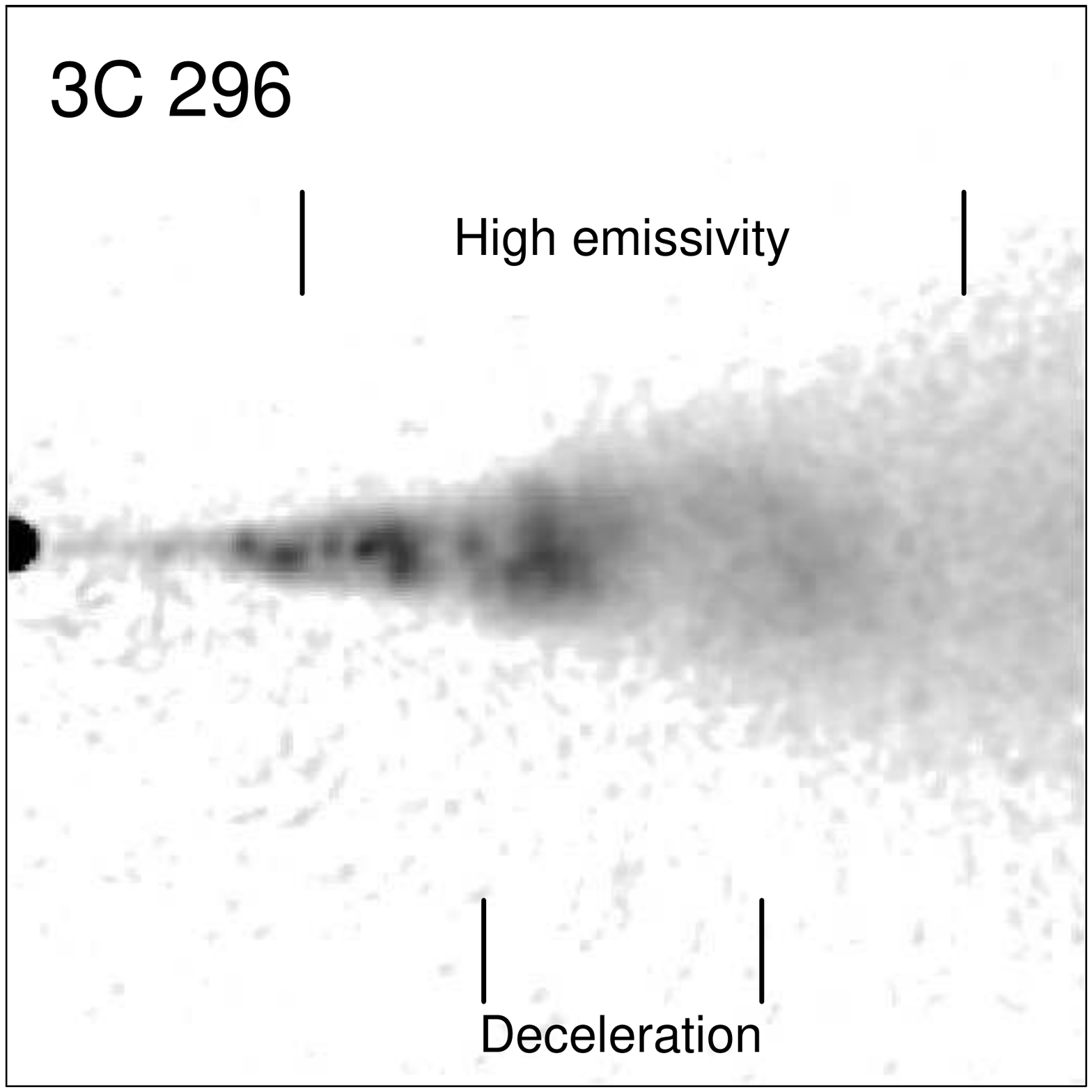} \\
\includegraphics[trim={0cm 2cm 0cm 3.cm},clip,width=0.45\textwidth]{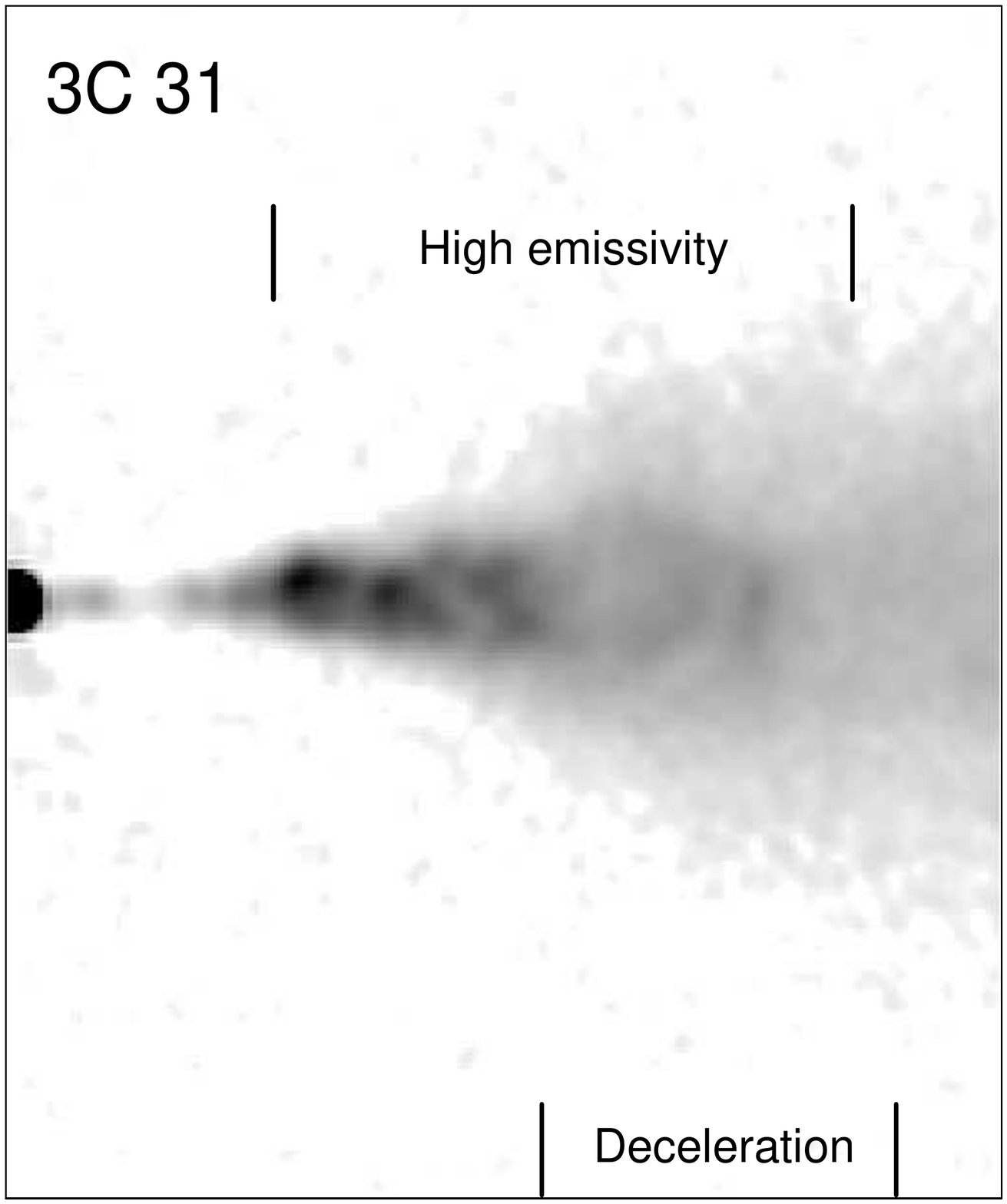} \quad \includegraphics[trim={0cm 3cm 0cm 4cm},clip,width=0.45\textwidth]{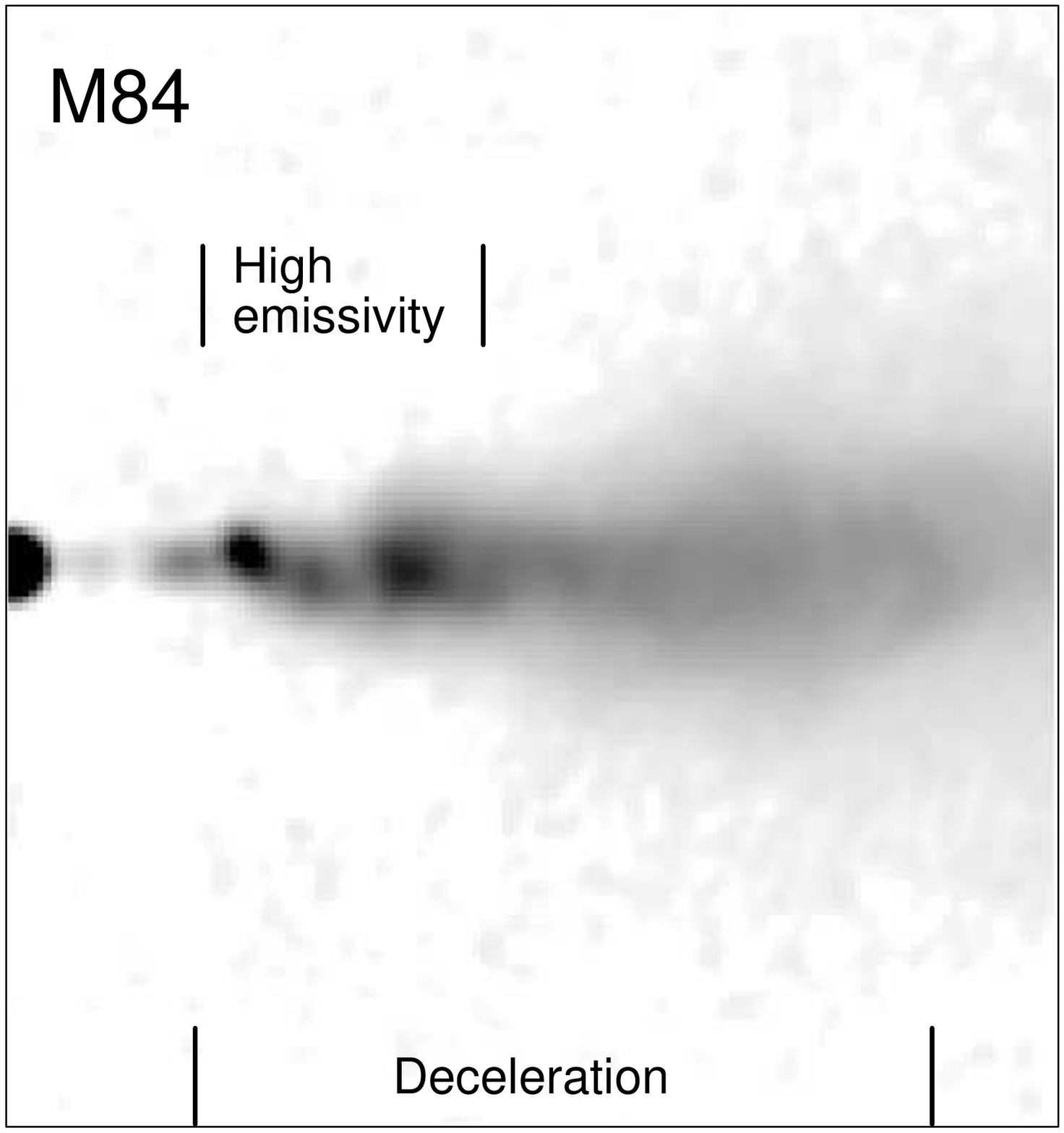}
\caption{Grey-scale plots of intensity (VLA) for the inner regions of NGC 315 (top left), 3C 296 (top right), 3C 31 (bottom left) and M84 (bottom right). The image shows the coincidence of the deceleration region (as obtained from detailed modeling) and those of high-emissivity \cite{lb14} (courtesy of Robert A. Laing).}\label{fig9}
\end{figure}

\section{Discussion}

 \subsection{Mass-load and dissipation}
 
     Bicknell \cite{bi84} and Komissarov \cite{ko90a} related the dissipation of kinetic energy to turbulence. Later, Bowman et al. \cite{bo96} showed that entrainment could cause a drop in kinetic energy along with an increase of internal energy. Figures~\ref{fig6} and \ref{fig9} show the relation between the deceleration region of FRI jets and diffuse emissivity that can be attached to dissipation of kinetic energy. In this section, I discuss this idea in the context of the FRI/FRII dichotomy and the production of VHE radiation in radio galaxies. The conservation equations \ref{eq:mass}, \ref{eq:mom} and \ref{eq:ene} that describe the evolution of a relativistic magnetized flow can be simplified, in the case of a steady-state, axisymmetric, and entraining jet propagating out of the galactic gravitational pull, with $v^\phi=v^r=0$ and $B^r=0$, to:
     
     

     \begin{equation} 
     \begin{split}
     \partial_z (\gamma \rho v^z)\, =\, q,\\
     \partial_z (\gamma \rho h^* v^z v^z + p^* - b^z b^z) \,=\, g^z, \\
     \partial_z (\rho h^* \gamma^2 v^z - b^0 b^z - \gamma \rho v^z) \,=\,v^z g^z,
     \end{split}
     \end{equation}    
where $\partial_z$ is the partial derivative with respect to the axial coordinate, $q$ is the mass-load per unit volume and per unit area and time, and $g^z$ is the gravitational acceleration. 
     
   If we focus on the energy conservation along a portion of the jet with parameters $\rho$, $v^z$, $\gamma$, and $B^\phi$ at a given position $z_0$, write the terms involving the magnetic field in the observer's frame, neglect the gravitational pull and discretize the equation, we find that the changes in the flow are given by:
   
   \begin{equation}
   \frac{\Delta \left[(\rho \gamma v^z) (h\gamma-1) R_j^2\right]}{\Delta z} +  \frac{\Delta \left[(B^\phi)^2 v^z R_j^2\right]}{\Delta z} \,=\,0.
   \end{equation}
   
  Taking into account, from the mass-conservation equation, that $\Delta (\rho \gamma v^z R_j ^2)/ \Delta z=q R_j^2$ (for a small opening angle), and neglecting the transfer of magnetic energy flux (once the jet expansion is conical this can only occur via magnetic dissipation processes, which, as stated above, are out of the ideal MHD approximation; see, e.g., \cite{ko12} and references therein), we can drop the magnetic term from the equation and obtain:
  
   \begin{equation}
   (\rho \gamma v^z)_0 \frac{\Delta (h\gamma)}{\Delta z} = (1 - (h\gamma)_0) q \,
   \end{equation}
which can be rewritten as
  
  
   \begin{equation}
   \frac{\Delta h}{h_0}=\frac{q \Delta z}{(\rho \gamma v^z)_0} \left(\frac{1}{(h\gamma)_0}-1\right) \,-\, \frac{\Delta \gamma}{\gamma_0}.
   \end{equation}
  
   This expression gives us an idea about the changes in the jet enthalpy as the jet flow evolves. We can now discuss the jet evolution in terms of $q / (\rho \gamma v^z)_0$, taking into account that $ (\Delta \gamma)/\gamma_0$ is small ($\sim 1$), and that $q / (\gamma \rho v^z)_0$ can be arbitrarily large (see also \cite{bo96}):
   

   \begin{enumerate}
   \item Strong relative mass-load ($q \Delta z \gg(\gamma \rho v^z)_0$): In this case $\Delta h < 0$ and the conservation equation tells us that the initial jet enthalpy is transferred to the entrained flow. 
   \item Mild relative mass-load ($q \Delta z \sim (\gamma \rho v^z)_0$): In this case $\Delta h$ could be both smaller than or larger than zero, depending on the value of the terms accounting for deceleration (so the initial jet enthalpy can grow). 
   \item Small relative mass-load ($q \Delta z \ll (\gamma \rho v^z)_0$): In this case, we could neglect the source term $q$ in the conservation equation above, and would be left with the Bernoulli expression for adiabatic evolution: $h\gamma = constant$, where expansion of a hot jet flow translates into acceleration.
   \end{enumerate}

   The third case could correspond to powerful jets, in which $q \Delta z \ll (\gamma \rho v^z)_0$, i.e., this is probably the case for FRIIs, as shown by numerical simulations of mass-entrainment by stellar winds in jets with different power \cite{pmlh14}, where it was reported that relatively weakly loaded jets do not suffer relevant dynamical effects (see Fig.~\ref{fig10}). In this case, dissipation of kinetic energy can occur at shocks (either stationary recollimation shocks or at traveling shocks) and the amount of dissipated energy (which is ultimately related to the emitted radiation) is regulated by the strength of these shocks, or by the growth of instabilities. In the case of recollimation shocks, these are stronger for hotter and slower flows, because the jet opening angle (determining the obliquity of the shock) increases with enthalpy and decreases with the Lorentz factor. Case 3 could thus be the scenario in FRII, where dissipation of kinetic energy occurs mainly at stationary shocks along the jet and at the interaction with the ambient medium (hotspot), close to which instabilities may have developed (e.g., Cygnus A).  
   
   
 
   
    Case 2 is a plausible mechanism for kinetic energy dissipation as long as the Lorentz factor is large enough to provide a significant contribution (in the terms that include $\Delta \gamma$). We could expect this heating process to play a role at the beginning of jet deceleration in FRIs (\cite{bo96}, Perucho et al., in preparation). Furthermore, the extra heating provided by mass-load naturally contributes to jet expansion, as observed in FRI jets from the so-called {\em flaring} point (e.g., \cite{lb02a}). 
      
\begin{figure}
\centering
\includegraphics[trim={0cm 9cm 0cm 9cm},clip,width=0.9\textwidth]{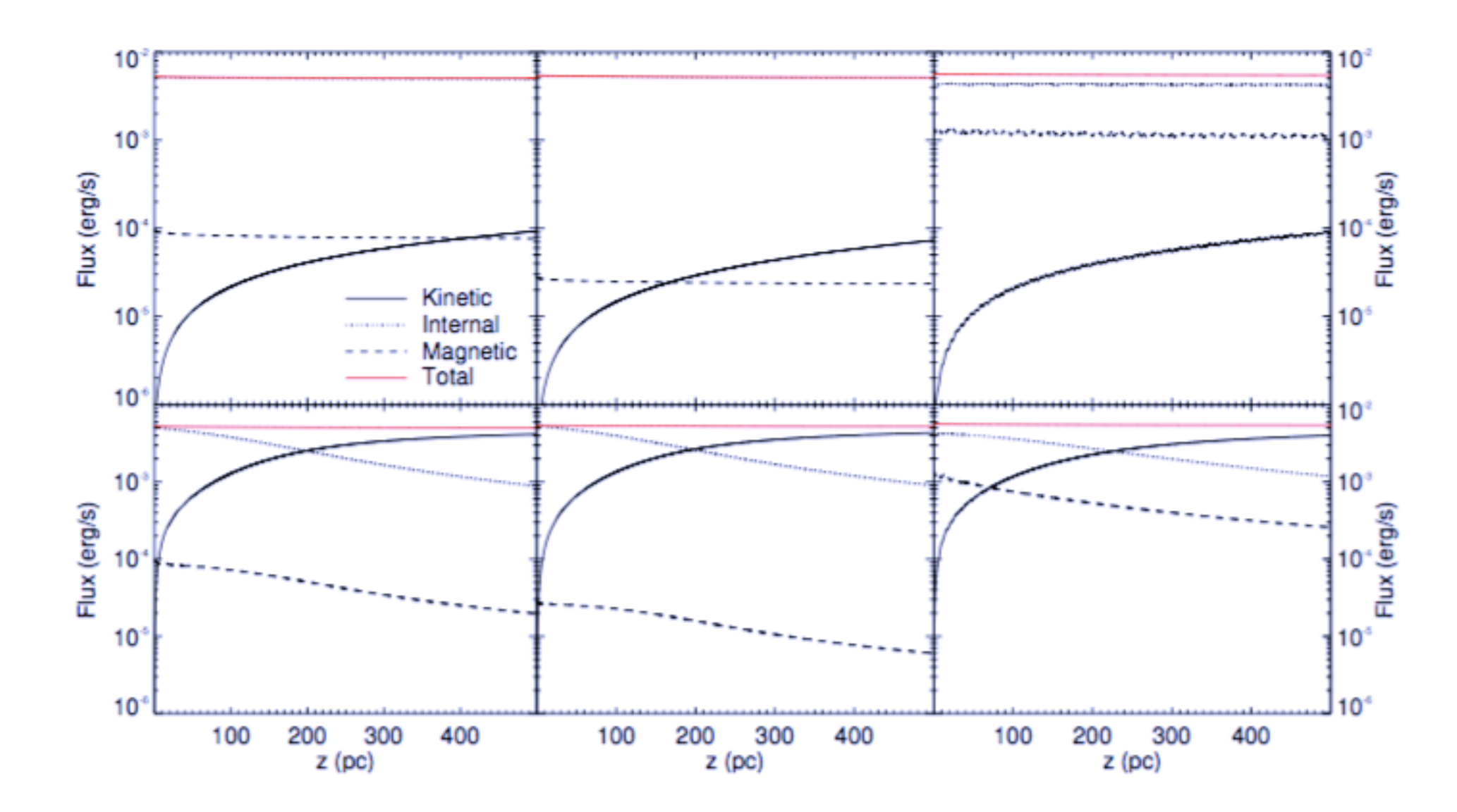}
\caption{Evolution of the energy fluxes (in logarithmic scale) for steady-state, axisymmetric jet simulations for hot dilute, magnetized jets mildly (mean stellar mass-loss $\dot{M}_w = 10^{-11}\,{\rm M_\odot/yr}$, upper panels) and strongly mass-loaded ($\dot{M}_w = 10^{-9}\,{\rm M_\odot/yr}$, lower panels). The jets also differ in the initial Lorentz factors 10 (left and right) and 6 (centre), and mean pitch angle (defined as $\tan^{-1}(B^\phi/B^z)$) of $45^{\circ}$ (left and centre) and $75^{\circ}$ (right). The red line indicates the total flux, the dotted line stands the internal energy flux, the solid line the kinetic energy flux and the dashed line the magnetic energy flux. Perucho et al. (in preparation).}\label{fig11}
\end{figure}

    In case 1, there is no possible gain of internal energy (via entrainment). We have to take into account that the jet is probably hot if the mass flux is small before entrainment begins ($(\rho \gamma v)_0 \ll$). As a result of strong relative entrainment in a hot, dilute jet, it is rapidly cooled and decelerated (this is also derived from simulations, \cite{bo96,pe14}, Perucho et al., in preparation). Dissipation and particle acceleration must thus be triggered by turbulence and locally, at shocks. The question is whether the role played by radiative output through this fast cooling is enough to break the assumption of adiabatic evolution or not. It has been actually reported that the flaring region in FRI jets differs from an adiabatic behavior \cite{lb14}. Figure~\ref{fig11} shows the evolution of the energy fluxes (in logarithmic scale) for steady-state, axisymmetric jet simulations for hot dilute, magnetized jets mildly and strongly entrained (from Perucho et al., in preparation). These simulations show that entrainment always implies an increase of kinetic energy flux in jets that have a very small percentage of kinetic energy flux at injection (i.e., hot and/or strongly magnetized) despite the deceleration, because of the increase of rest-mass density. Depending on the relative role of entrainment, the kinetic energy can grow at the expense of internal energy flux alone (relatively mild entrainment), or at the expense of both internal and magnetic energy fluxes (strong entrainment). The authors reach the conclusion that jets are not heavily affected by mass-load as long as the initial mass flux per unit area at $z_0$ (with $v^z \simeq c$):
    
  \begin{eqnarray} \label{eq:mcrit}
  \rho_{j,0} \,\gamma_{j,0} \, c > \,6.7 \times 10^{-31}\, \left(\frac{\dot{M}}{10^{-12} {\rm M_\odot}{\rm yr}^{-1}}\right)\, \, 
\left(\frac{n_s}{10\,{\rm pc^{-3}}}\right) \, \left(\frac{\Delta z}{1\, {\rm kpc}}\right)^3 \left( \frac{\tan(\alpha)}{\tan(1^\circ)}\right)^2 \left(\frac{R_{j,0}}{1\, {\rm pc}}\right)^{-2}  \, {\rm g\,cm^{-3}},
  \end{eqnarray}
where $\alpha$ is the jet opening angle. This expression does not take into account the role of magnetic tension, heating by dissipation of kinetic energy, the sudden increase of the jet opening angle once mass-loading starts to be relevant (because of the increased pressure), or the drop in stellar density along $\Delta z$. Nevertheless, it represents a good a priori estimate of the expected role that stellar mass-load can play on jet dynamics for given initial conditions.
    
\begin{figure}
\centering
\includegraphics[trim={0cm 0cm 0cm 1cm},clip,width=\textwidth]{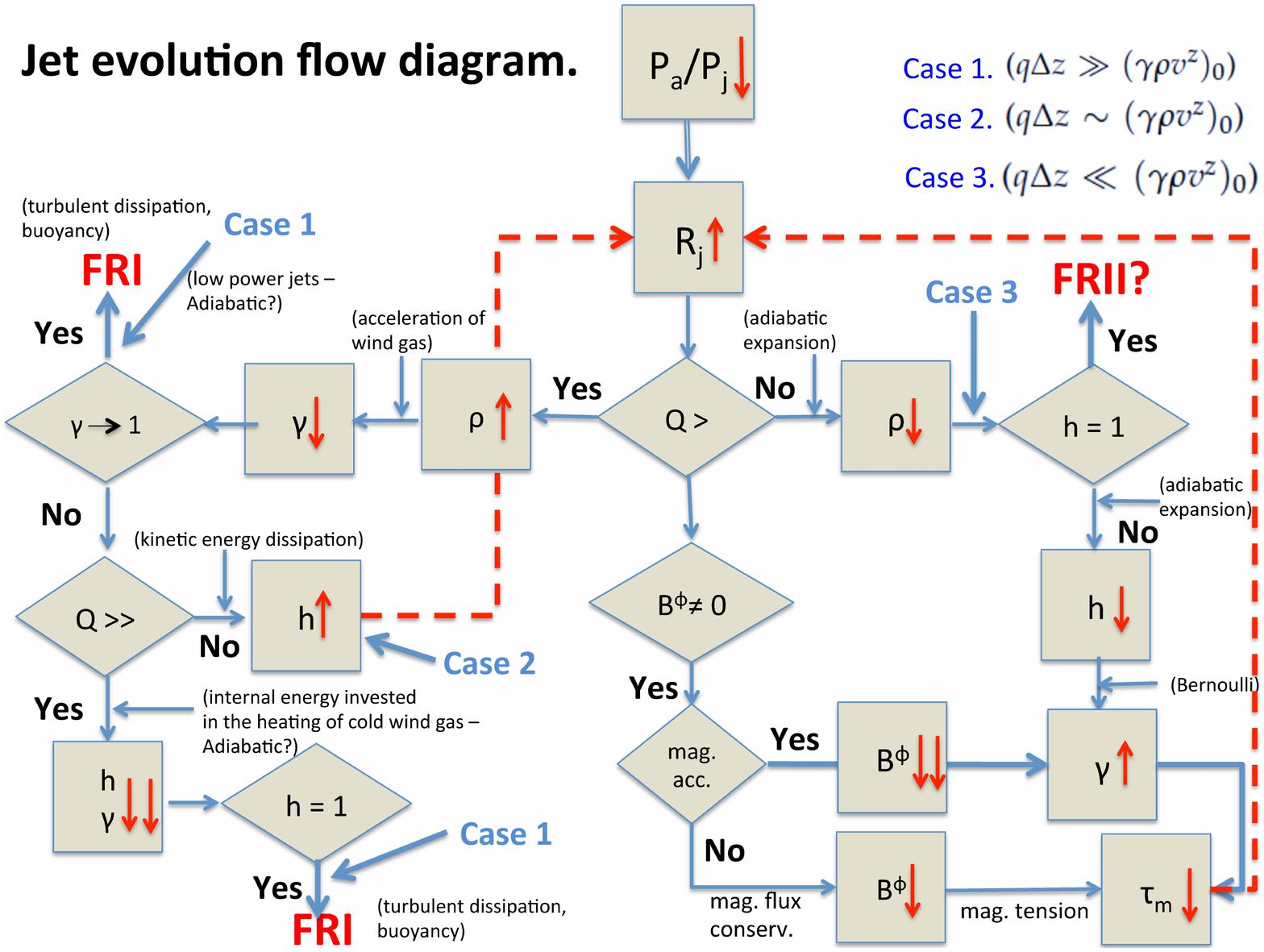}
\caption{Flow diagram of jet evolution as it propagates through a galactic atmosphere. The rhombi indicate the bifurcations in the flow diagram depending on which processes dominate the dynamics, and the squares indicate the result of those processes. I have indicated the possible evolutionary options related to mass load in blue. Adapted from Perucho et al., in preparation (see also \cite{pe19b}). Note that where case 1 is indicated, the jet evolution can be non-adiabatic \cite{lb14}. Case 2 may imply particle acceleration and enhanced radiative output, still keeping an adiabatic behavior of the flow.}\label{fig12}
\end{figure}

    Both for cases 1 and 2, a cooled and decelerated jet leads to $h \gamma \rightarrow 1$, with the initial jet luminosity having been invested into kinetic energy flux (dominated by the entrained particles) and a part having been radiated away. In other words, the process of entrainment and deceleration leads to the depletion of the relativistic reservoir of the jet from parsec to kiloparsec scales. Once the jet becomes decelerated to subrelativistic speeds, the classical equations take over the description of jet evolution and the appropriate approach then becomes the one described in \cite{bi84} and/or \cite{ko90a}, including relevant terms such as buoyant forces and turbulent dissipation fed by further entrainment. Figure~\ref{fig12} shows a flow diagram of expanding jets, where I have indicated the processes that would correspond to the three mass-loading cases described here. Note that this is independent from the origin of the entrainment. When following the diagram, it has to be kept in mind that different processes can take place at the same time.   
    
    Finally, it is important to stress that, independently from the dynamical role of entrainment, the entrainment of protons in the jet can significantly change the jet composition along its propagation.

\subsection{Jet evolution: a summary}

  Jets are probably magnetically dominated at the formation region, with their mass flux dominated by pairs \cite{bz77}. Magnetic acceleration and the Bernoulli effect convert part of the magnetic flux and internal energy into kinetic energy from the inner parsec to possibly hundreds of parsecs \cite{ho15}. After the initial injection carves its way through the irregular inner galactic ambient medium, a channel forms that allows the particles to propagate at relativistic speeds along it. The CDI and KHI are assumed to develop at different scales along the jet, with the former acting while the jet is magnetically dominated and the KHI downstream of the magnetic acceleration region. The CDI (including the pressure-driven instability) could actually play a significant role in dissipating magnetic energy and trigger particle acceleration to relativistic energies, feeding the non-thermal population. As the jets expand in the external density/pressure gradients, the large-scale instabilities can be stabilized (see Section~\ref{sec:nlr}).
    
   In parallel, as the jet flow propagates through the host galaxy, the interaction with stars and clouds can mass-load the jet, introducing protons and forcing the jet deceleration and the conversion of internal and/or magnetic energy flux into kinetic energy. In the case of fast jets, the deceleration caused by mass-load can produce an increase in the internal energy flux (e.g., \cite{bo96,pmlh14}, Perucho et al., in preparation). Deceleration of jets by stellar winds has been proven to be efficient in low-power jets, even for old stellar populations with weak winds \cite{pmlh14}. Jet/obstacle interactions also probably generate strong pressure and velocity inhomogeneities across the jet.  
   
   The expansion of a supersonic flow is followed by recollimation, a process that can trigger the development of RTI/CFI. Taking into account that these instabilities are enhanced in the case of larger opening angles \cite{gk18a,gk18b}, we expect them to develop in slower and hotter jets (i.e, jets with smaller magnetosonic Mach numbers). 
   
      Although such short wavelength instabilities are favoured by the progressive deceleration that starts at the jet boundaries \cite{lb14}, other mechanisms can act in parallel, contributing to the FRI jet disruption process. Most FRI jets show a brightening region both in radio and X-rays, and, in some of them (see Fig.~\ref{fig9} and, in particular, the case of NGC 315, discussed in \cite{wo07}), there are clear hints of large-amplitude (i.e., probably disruptive) helical structures within the broader conically expanding jet, beyond the brightening region. Whether these helices are revealing the development of instabilities in the inner jet spine embedded within a broadening mixing layer is unclear, but it remains a plausible option. The symmetric brightening of the jet and counter-jet, previous to geometrical flaring could also be related a recollimation shock (e.g., \cite{pm07}). The detection of high-energy emission from large-scales in FRIs requires that the decelerating process is accompanied by particle acceleration, and turbulence seems to be a good candidate to explain this.
   
   Kinetically dominated jets (fast and cold) have smaller opening angles and dissipate kinetic energy mainly at the recollimation shocks and at their hotspots, but also possibly due to the development of small-scale instabilities in the boundary layer and at jet-obstacle interactions (see, e.g., \cite{ha16} on Pictor A, where the X-rays have been proposed to be synchrotron emission from the jet boundary layer). Although we do not observe clear hints of large-scale jet disruption at kpc-scales, we do observe helical structures (probably coupled to KHI modes) indicating that dissipation can take place both along the jet flow and at its head (via the dentist drill effect, \cite{sch82}, \cite{pe12,pe12b,vg19}).
      
      Recent simulations show that the propagation of FRII jets to hundreds of kiloparsecs can take place in 1-10~Myr, aided by the development through a decreasing density atmosphere and small scale oscillations of the jet head \cite{pe19}, until the oscillations grow to nonlinear amplitudes and then the dentist drill mechanism comes into play, decelerating the advance of the jet head. This could represent a transitory phase in the evolution of some FRII jets, most probably in those showing thin lobes and remarkable collimation. Instabilities would take long enough to grow to allow for large-scale jet propagation, and would start to contribute to strong kinetic energy dissipation only at those scales. Finally, from the lack of strong jet deceleration to subrelativistic speeds in FRII jets, we can infer that entrainment does not play a significant role in their evolution.

\subsection{A final comment on FRI/FRII dichotomy and VHE emission}         
         
    A conclusion that can be derived from this discussion is that dissipation is not acting as strongly in FRII jets as in FRIs at large scales. The dissipation in FRIs happens via processes that can significantly contribute to particle acceleration (e.g., strong shocks, shear, and turbulent mixing) and jet deceleration. Furthermore, larger opening angles in FRI jets after the flaring (which coincides with the beginning of deceleration, see, e.g., \cite{lb02a,lb14}) contribute to rapidly increase the entrainment rates and also the number of stars embedded in the jet, thus favoring further investment of jet kinetic flux into particle acceleration at interaction sites (e.g., \cite{vi17}). Finally, deceleration can reduce the Doppler boosting and favor the detection in misaligned sources. In contrast, in more powerful FRII jets the number of interactions is necessarily smaller because they are more collimated. Furthermore, the radiation that they trigger at interactions can be boosted far from our view. Altogether, this could contribute to explain why FRI radio galaxies are more numerous in the Fermi catalog than FRIIs, an observational fact that cannot be explained by means of simple population numbers (e.g., \cite{gr12}). 
    
    The previous argumentation is valid for the diffuse HE radiation from FRI jets, and mass-loading has actually been successfully used to reproduce the broad band synchrotron spectrum of radio galaxies from radio to X-rays, such as Centaurus~A \cite{wyk15}. Actually, evidence is being found that shows the lack of correlation between Doppler boosting and gamma-ray emission in radiogalaxies (\cite{an19} and references therein), as opposed to blazars (see also \cite{ar12,boe16}), where the fast variability of the detected VHE radiation seems to imply that it is generated at the most compact scales, via IC or even proton-proton interaction (see \cite{rl18} for a recent review). This points towards a possible decoupling of the gamma-ray emission from radiogalaxies and their parsec-scale jets.
        
    In this review I have summarized candidate scenarios for energy dissipation and particle acceleration in radio galaxies. These scenarios are probably acting more efficiently in FRI jets, but the exact way in which TeV photons are produced in these sources remains to be determined, although it could be related to 1) enhanced efficiency (with respect to our current acceleration models) of particle acceleration in any of the aforementioned scenarios (via acceleration in turbulent flows), or 2) processes not taken into account in ideal RMHD, such as reconnection caused by the development of instabilities or jet-star/cloud interactions. The detailed multi-wavelength observational study and modeling of the dynamics of FRI radiogalaxies (e.g., \cite{lb14,pmlh14}, Perucho et al., in preparation) and nearby low-power, gamma-ray bright active galaxies like Centaurus~A or NGC~1052 (e.g., \cite{wo08,ab10,go10,hc11,mu14,fr18,ba19,fr19}), plus particle-in-cell simulations (see, e.g., \cite{ni19,ma19,wo19}, and references therein), set the path to understanding the relation between jet deceleration and high-energy emission.




\vspace{6pt} 




\funding{MP has been supported by the Spanish Ministerio de Ciencia y
Universidades (grants AYA2015-66899-C2-1-P and AYA2016-77237-C3-3-P)
and the Generalitat Valenciana (grant PROMETEOII/2014/069).}

\acknowledgments{I thank Jos\'e Mar\'{\i}a Mart\'{\i} and Valent\'{\i} Bosch-Ramon for the discussions that have contributed to the ideas that I expose in this review, and for comments to the original manuscript. I thank J. Matsumoto, K. Gourgouliatos, G. Bodo, M. Hardcastle, S. Komissarov, and R. Laing for giving permission to add figures for this publication. I also thank the graduate student N\'uria Chiquillo for providing tex equations that have eased the writing of part of the text.}

\conflictsofinterest{The authors declare no conflict of interest. The funders had no role in the design of the study; in the collection, analyses, or interpretation of data; in the writing of the manuscript, or in the decision to publish the results.} 

\reftitle{References}





\end{document}